\documentclass[sigconf, screen]{acmart}
\AtBeginDocument{%
  }

\usepackage{graphicx}
\usepackage{booktabs}
\usepackage{multirow}
\usepackage{amsmath}

\usepackage[most]{tcolorbox}
\usepackage{xcolor}
\usepackage{enumitem}
\usepackage{setspace}

\usepackage[table]{xcolor}
\usepackage{colortbl}
\usepackage{subcaption}
\usepackage{enumitem}
\usepackage{algorithm}
\usepackage{algorithmic}
\usepackage{pifont}
\usepackage{makecell}
\usepackage{xspace}

\newcommand{\modelname}{\text{AVID}\xspace}
\newcommand{\baselinename}{\text{AVID-Qwen}\xspace}

\setcopyright{acmlicensed}
\copyrightyear{2026}
\acmYear{2026}
\acmDOI{XXXXXXX.XXXXXXX}
\acmConference[ACM MM '26]{Proceedings of the 34th ACM International Conference on Multimedia}{October 2026}{TBD}
\acmISBN{978-1-4503-XXXX-X/26/10}

\begin{document}

\title{AVID: A Benchmark for Omni-Modal Audio-Visual Inconsistency Understanding via Agent-Driven Construction}

\author{Zixuan Chen}
\affiliation{\institution{Shanghai Jiao Tong University}\country{China}}
\email{13924560444@sjtu.edu.cn}

\author{Depeng Wang}
\affiliation{\institution{Ant Group}\country{China}}

\author{Hao Lin}
\affiliation{\institution{The Chinese University of Hong Kong}\country{China}}

\author{Li Luo}
\affiliation{\institution{Shanghai Jiao Tong University}\country{China}}

\author{Ke Xu}
\affiliation{\institution{Shanghai Jiao Tong University}\country{China}}

\author{Ya Guo}
\authornote{Corresponding author}
\affiliation{\institution{Ant Group}\country{China}}

\author{Huijia Zhu}
\affiliation{\institution{Ant Group}\country{China}}

\author{Tanfeng Sun}
\affiliation{\institution{Shanghai Jiao Tong University}\country{China}}

\author{Xinghao Jiang}
\authornote{Corresponding author}
\affiliation{\institution{Shanghai Jiao Tong University}\country{China}}

\date{}

\begin{abstract}
We present AVID, the first large-scale benchmark for audio-visual inconsistency understanding in videos. While omni-modal large language models excel at temporally aligned tasks such as captioning and question answering, they struggle to perceive cross-modal conflicts,a fundamental human capability that is critical for trustworthy AI. Existing benchmarks predominantly focus on aligned events or deepfake detection, leaving a significant gap in evaluating inconsistency perception in long-form video contexts.
AVID addresses this with: (1) a scalable construction pipeline comprising temporal segmentation that classifies video content into Active Speaker, Voiceover, and Scenic categories; an agent-driven strategy planner that selects semantically appropriate inconsistency categories; and five specialized injectors for diverse audio-visual conflict injection; (2) 11.2K long videos (avg. 235.5s) with 39.4K annotated inconsistency events and 78.7K segment clips, supporting evaluation across detection, temporal grounding, classification, and reasoning with 8 fine-grained inconsistency categories.
Comprehensive evaluations of state-of-the-art omni-models reveal significant limitations in temporal grounding and reasoning. Our fine-tuned baseline, AVID-Qwen, achieves substantial improvements over the base model (2.8$\times$ higher BLEU-4 in segment reasoning) and surpasses all compared models in temporal grounding (mIoU: 36.1\% vs 26.2\%) and holistic understanding (SODA-m: 7.47 vs 6.15), validating AVID as an effective testbed for advancing trustworthy omni-modal AI systems.\url{https://github.com/czx1220/AVID-bench}
\end{abstract}

\ccsdesc[500]{Computing methodologies~Video understanding}
\ccsdesc[300]{Computing methodologies~Natural language processing}
\keywords{Audio-Visual Inconsistency, Multimodal Benchmark, Temporal Grounding, Omni-modal Understanding}

\ccsdesc[500]{Computing methodologies~Video understanding}
\ccsdesc[300]{Computing methodologies~Natural language processing}

\keywords{Audio-Visual Inconsistency, Multimodal Benchmark, Temporal Grounding, Omni-modal Understanding}

\begin{teaserfigure}
    \centering
    \includegraphics[width=1\linewidth]{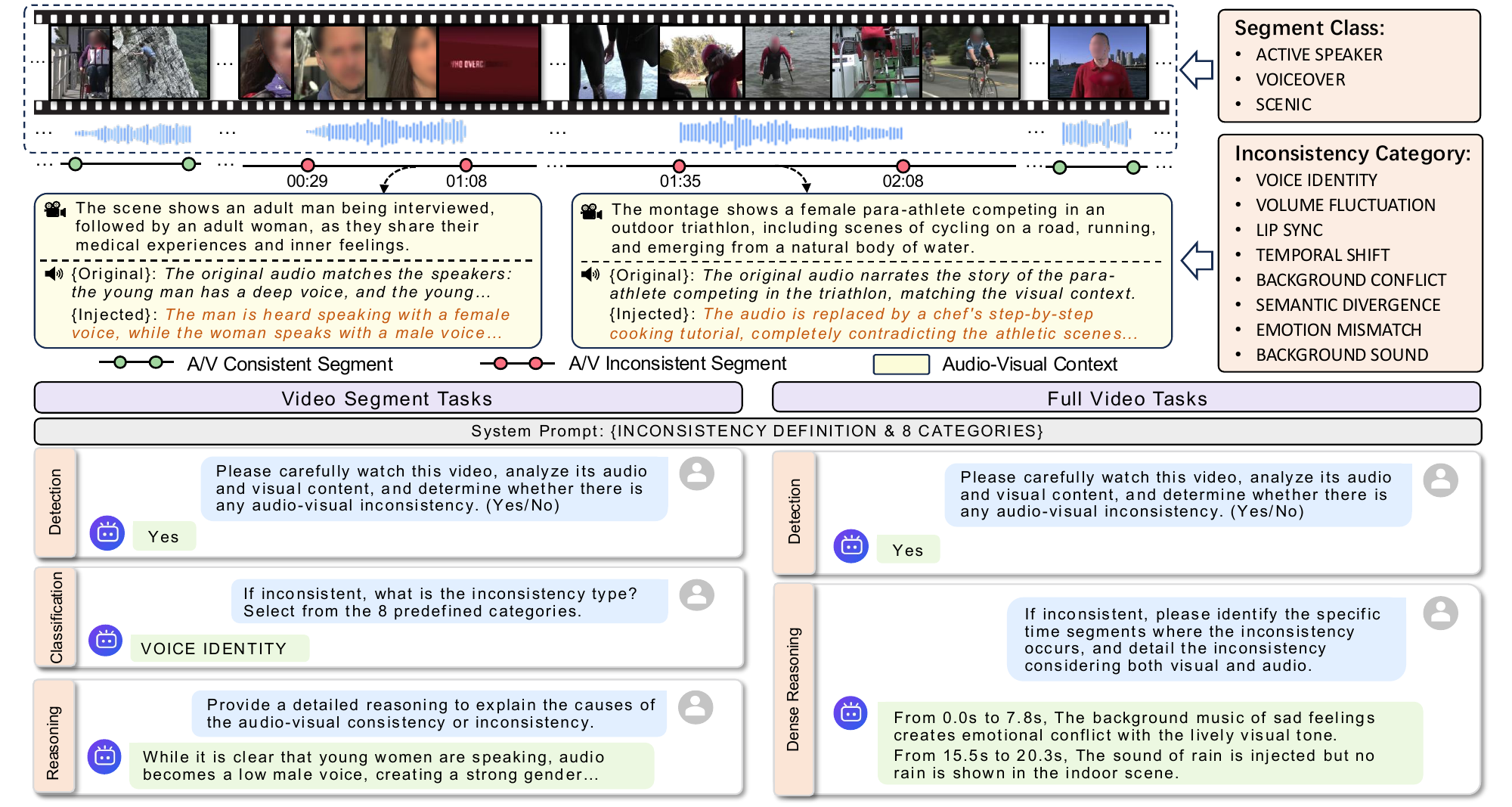}
    \caption{We introduce AVID, the first benchmark with explicit taxonomy and formal definitions for audio-visual inconsistency understanding. AVID comprises two complementary evaluation settings: (1) a segment-level set of 78K+ clips, each either fully consistent or fully inconsistent, supporting inconsistency detection, category classification, and fine-grained reasoning; and (2) a full-video set of 11K+ full videos containing both consistent and inconsistent segments, supporting inconsistency detection, temporal grounding, and dense reasoning. All samples are annotated with precise temporal boundaries, 8 inconsistency categories across 3 segment classes, and detailed causal explanations.
    }
    \label{fig:bench_overview}
\end{teaserfigure}

\maketitle

\section{Introduction}\label{sec:intro}

The emergence of omni-modal large language models, such as Gemini 3.1 Pro, MiMo-V2-Omni, and Qwen3-Omni, has significantly advanced machine perception. By jointly modeling visual, auditory, and textual signals, these models have achieved strong performance on tasks such as video captioning, visual question answering (VQA), and temporal action localization~\cite{geng2025longvale, goel2024omcat,videollama,fu2025video}. However, one fundamental cognitive ability remains largely overlooked in standard evaluation protocols: the understanding of audio-visual \emph{inconsistency}~\cite{ye2026eyes}. Human perception involves not only recognizing objects and sounds, but also validating whether they are physically, semantically, and spatially coherent~\cite{um2025object}. For instance, humans can immediately notice when an elderly man speaks with the voice of a young woman, or when the audio description conflicts with the visual content, as shown in Figure~\ref{fig:bench_overview}.

Despite its importance for trustworthy AI, hallucination detection, deepfake mitigation, and enhanced audio-visual understanding~\cite{huang2024trustllm,wang2023amber}, this capability is not comprehensively evaluated by current multi-modal video benchmarks. We identify two major limitations in existing datasets:  
\textbf{Limitation 1: Positive-only bias in grounding benchmarks.} Mainstream audio-visual grounding benchmarks, such as AVE~\cite{wu2019dual}, UnAV-100~\cite{geng2023dense}, and LongVALE~\cite{geng2025longvale}, focus exclusively on localizing temporally aligned positive events. They implicitly assume that visual and auditory streams are always congruent, making them unsuitable for evaluating models' robustness against hallucinated or tampered multi-modal content~\cite{chen2020vggsound,gemmeke2017audio}.  
\textbf{Limitation 2: Artifact-oriented bias in deepfake datasets.} Deepfake and synchronization datasets, such as FakeAVCeleb~\cite{yang_avoid-df_2023}, LAV-DF~\cite{cai_you_2022}, and AV-Deepfake1M++~\cite{cai_av-deepfake1m_2025}, aim to address audio-visual manipulations. However, models can often achieve high accuracy by exploiting uni-modal artifacts—such as visual blending defects or unnatural audio frequencies—without genuinely understanding the cross-modal correlations~\cite{xie_deepfake_2026,haliassos2021lips,li2020face}. Furthermore, these manipulations are typically confined to extremely short durations~\cite{liu_lips_2024, bohacek_lost_2024}, which are insufficient for assessing anomaly perception in long-form video contexts. As summarized in Table~\ref{tab:comparison}, these benchmarks fail to fully address the broader challenges posed by real-world audio-visual inconsistencies.

To address these limitations, we introduce \textbf{AVID}, the first benchmark specifically designed to systematically evaluate audio-visual inconsistency understanding in videos. We develop a comprehensive and scalable framework for generating inconsistent video segments, which consists of three key stages. First, due to the varying types of inconsistencies across different video segments, we segment and classify videos into \textbf{ACTIVE SPEAKER}, \textbf{VOICEOVER}, and \textbf{SCENIC} categories using speaker detection and voice activity analysis. Second, a strategy agent analyzes the content of each segment and recommends suitable types of inconsistency to inject. Third, guided by the strategy agent's recommendations, we apply five injectors to introduce diverse and controlled audio-visual conflicts while preserving the original visual content and video duration.

Building on this framework, we construct \textbf{AVID}, a full-video benchmark for audio-visual inconsistency understanding. As shown in Table~\ref{tab:comparison}, AVID includes \textbf{11.2K} full videos, with a 3:1 ratio between videos containing inconsistency events and fully consistent videos. It also includes \textbf{39.4K} annotated inconsistent event segments, each with high-quality temporal boundaries and detailed reasoning about the underlying audio-visual conflict. Additionally, we create a segment-level subset with \textbf{78.7K} clips, ensuring a balanced positive-to-negative ratio (1:1). The average duration for full videos is \textbf{235.5s}, and for segment clips, it is \textbf{14.8s}. AVID supports evaluation at both the segment level and the full-video level, as shown in Figure~\ref{fig:bench_overview}.

We also evaluate representative open-source and closed-source omni-modal models on AVID, conducting a fine-grained analysis of their ability to handle inconsistency tasks. Our results reveal clear limitations in existing systems, particularly in inconsistency detection, temporal localization, and fine-grained reasoning. Furthermore, we introduce \textbf{AVID-Qwen}, a Qwen3-Omni model fine-tuned on AVID, demonstrating that AVID significantly improves inconsistency perception and cross-modal reasoning capabilities. Our contributions are summarized as follows:

\begin{itemize}[leftmargin=*]
    \item \textbf{A novel data construction pipeline.} We design a scalable and semantically grounded framework for generating inconsistent video segments, which includes segment typing, strategy-guided inconsistency planning, and controlled inconsistency injection. This pipeline enables the creation of diverse inconsistency patterns while retaining the original visual content and temporal structure.
    \item \textbf{AVID: A benchmark for audio-visual inconsistency understanding.} We present the first benchmark tailored for evaluating audio-visual inconsistency understanding in videos. AVID contains \textbf{11.2K} full videos, \textbf{39.4K} inconsistent event segments, and \textbf{78.7K} video clips, with fine-grained category labels, temporal annotations, and reasoning about inconsistencies. AVID enables evaluation at both segment and full-video levels.
    \item \textbf{Comprehensive evaluation and a strong tuning baseline.} We evaluate leading open-source and closed-source omni-modal models on AVID and provide a fine-grained analysis of their performance in inconsistency detection, localization, and reasoning tasks. Additionally, we demonstrate that fine-tuning \textbf{AVID-Qwen} on AVID substantially improves inconsistency understanding and cross-modal reasoning.
\end{itemize}

\begin{table*}[t]
     \centering
     \caption{Comparison with representative audio-visual benchmarks from the perspective of inconsistency understanding.
     \textbf{Video Form}: segment-level clips or full-video samples.
     \textbf{A/V Inconsistency}: whether the benchmark explicitly targets audio-visual inconsistency.
     \textbf{\#Inc. Cat.}: number of inconsistency categories.
     \textbf{Event Grounding}: whether event-level temporal localization annotations are provided.
     \textbf{Inc. Reasoning}: whether judging inconsistency requires explicit reasoning.
     G: generated. M: manual. V: visual. A: audio. S: speech. Caption modality indicates which modality the caption is based on. ``--'' denotes not applicable for benchmarks that do not target audio-visual inconsistency.}
     \label{tab:comparison}
     \resizebox{\textwidth}{!}{%
     \begin{tabular}{l c c r r c c c c c}
     \toprule
     \textbf{Dataset} & \textbf{Anno.} & \textbf{Video Form} & \textbf{\#Videos} & \textbf{Avg.\,Len} & \textbf{Caption} & \textbf{A/V Inc.} &
     \textbf{\#Inc. Cat.} & \textbf{Inc. Reasoning} & \textbf{Timestamp} \\
     \midrule
     \multicolumn{10}{l}{\textit{Grounding / Localization Benchmarks}} \\
     \midrule
     InternVid~\cite{wang2023internvid}      & G   & Segment    & 234M  & 11.7\,s  & V     & \ding{55} & -- & -- & \ding{55} \\
     Panda-70M~\cite{chen2024panda}          & G   & Segment    & 70.8M & 8.5\,s   & V     & \ding{55} & -- & -- & \ding{55} \\
     AudioCaps~\cite{kim2019audiocaps}       & M   & Segment    & 39.6K & 10\,s    & A     & \ding{55} & -- & -- & \ding{55} \\
     WavCaps~\cite{mei2024wavcaps}           & G   & Segment    & 403K  & 67.6\,s  & A     & \ding{55} & -- & -- & \ding{55} \\
     ACAV100M~\cite{lee2021acav}             & G   & Segment    & 100M  & 10\,s    & --    & \ding{55} & -- & -- & \ding{55} \\
     VALOR~\cite{liu2024valor}               & M   & Segment    & 1.18M & 10\,s    & VA    & \ding{55} & -- & -- & \ding{55} \\
     VAST~\cite{chen2023vast}                & G   & Segment    & 27M   & --       & VAS   & \ding{55} & -- & -- & \ding{55} \\
     AVE~\cite{wu2019dual}                   & M   & Segment    & 4,143 & 10\,s    & VA    & \ding{55} & -- & -- & \ding{51} \\
     UnAV-100~\cite{geng2023dense}           & M   & Full Video & 10,790& 42.1\,s  & --    & \ding{55} & -- & -- & \ding{51} \\
     ActivityNet Caps~\cite{krishna2017anet} & M   & Full Video & 20K   & 180\,s   & V     & \ding{55} & -- & -- & \ding{51} \\
     Charades-STA~\cite{gao2017tall}         & G+M & Full Video & 10K   & 30\,s    & V     & \ding{55} & -- & -- & \ding{51} \\
     LongVALE~\cite{geng2025longvale}        & G+M & Full Video & 8,411 & 235\,s   & VAS   & \ding{55} & -- & -- & \ding{51} \\
     \midrule
     \multicolumn{10}{l}{\textit{Deepfake / Synchronization Benchmarks}} \\
     \midrule
     FakeAVCeleb~\cite{khalid2021fakeavceleb}       & G+M & Segment & 20,000 & 7.8\,s & -- & \ding{51} & 3 & \ding{55} & \ding{55} \\
     LAV-DF~\cite{cai_you_2022}                     & G   & Segment & 136K   & --     & -- & \ding{51} & 3 & \ding{55} & \ding{51} \\
     LipFD / AVLips~\cite{liu_lips_2024}            & G   & Segment & 340K   & --     & -- & \ding{51} & 1 & \ding{55} & \ding{55} \\
     AV-Deepfake1M++~\cite{cai_av-deepfake1m_2025}  & G   & Segment & 2M     & 8.2\,s & -- & \ding{51} & 3 & \ding{55} & \ding{51} \\
     \midrule
     \multicolumn{10}{l}{\textit{Audio-Visual Inconsistency Benchmark (Ours)}} \\
     \midrule
     \multirow{2}{*}{\textbf{AVID (Ours)}} & \multirow{2}{*}{G+M} & Segment    & \textbf{78,722} & \textbf{14.8s}  & \multirow{2}{*}{VAS} & \multirow{2}{*}{\textbf{\ding{51}}} & \multirow{2}{*}{\textbf{8}} & \multirow{2}{*}{\ding{51}} & \multirow{2}{*}{\ding{51}} \\
      &  & Full Video & \textbf{11200} & \textbf{235.5s} &  &  &  &  &  \\
     \bottomrule
     \end{tabular}%
     }
\end{table*}

\begin{figure*}[t]
    \centering
    \includegraphics[width=1\linewidth]{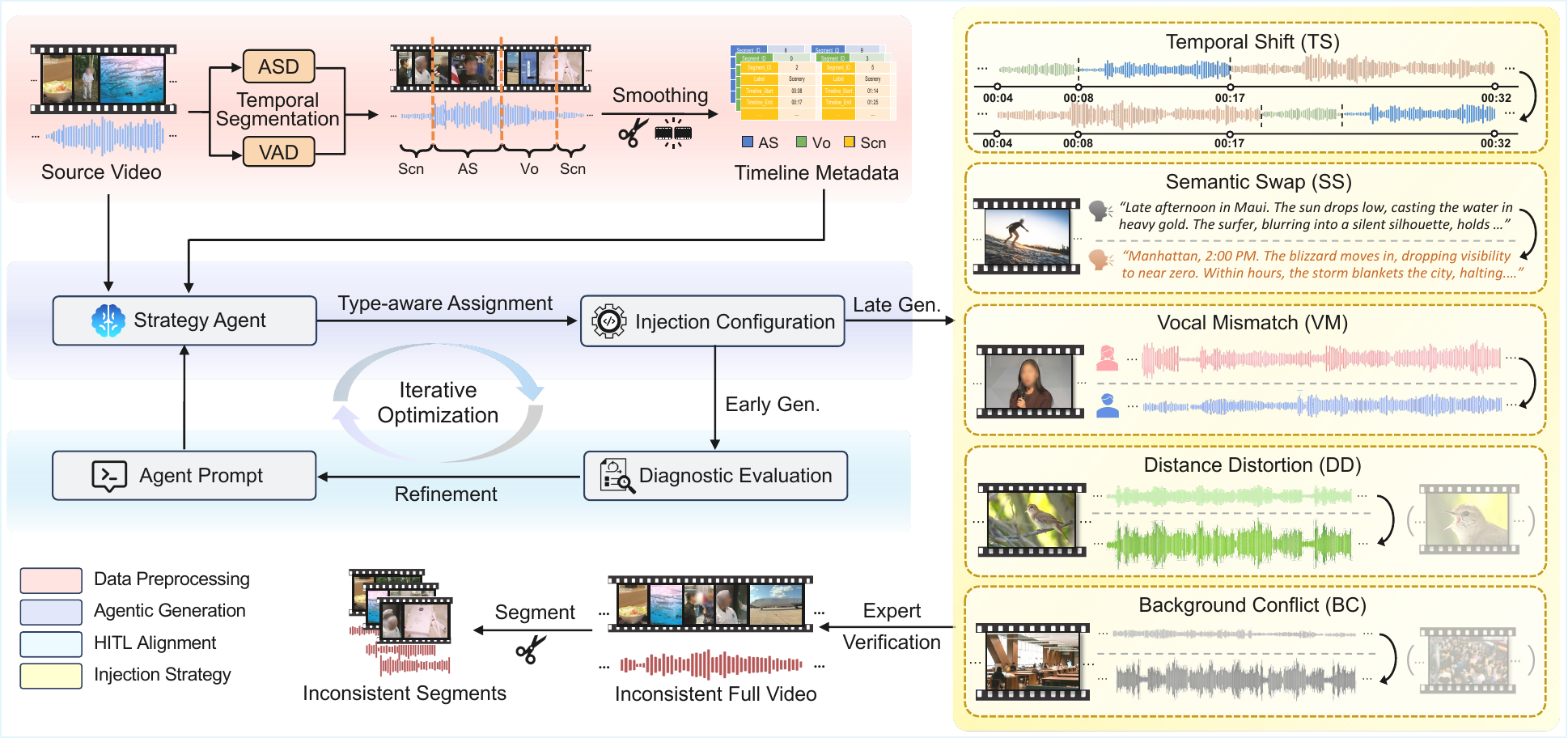}
    \caption{Overview of the AVID construction pipeline, consisting of three stages: (1) data preprocessing with temporal segmentation and class-aware labeling, (2) agentic generation with strategy planning and five specialized inconsistency injectors, and (3) human-in-the-loop alignment for quality control.}
    \label{fig:benchmark_pipeline2}
\end{figure*}

\section{Related Work}\label{sec:related}

\noindent\textbf{Omni Model Audio-Visual Benchmarking.}  
The joint processing of audio and visual signals has significantly advanced scene understanding~\cite{zhu2021deep,owens2018audio}. Early benchmarks such as AVE~\cite{wu2019dual} and LLP~\cite{tian2020unified} localized audio-visual events in untrimmed videos. More recent datasets, including UnAV-100~\cite{geng2023dense} and LongVALE~\cite{geng2025longvale}, introduced dense multi-event localization and minute-long temporal grounding. However, these datasets primarily focus on \textit{naturally occurring, aligned} events, assuming congruence between visual and auditory streams, and thus fail to evaluate models' robustness against hallucinated or tampered content.

\noindent\textbf{Deepfake and Inconsistency Detection.}
Detecting manipulated media is critical for security~\cite{tolosana2020deepfakes}. Traditional deepfake datasets like FakeAVCeleb~\cite{yang_avoid-df_2023} focus on facial manipulations and lip-sync errors, relying on single-modal artifact detection~\cite{aleem_seeing_2026} without requiring a multi-modal understanding of audio-visual inconsistencies. Recent works such as LAV-DF~\cite{cai_you_2022} and Liu et al.~\cite{liu_audio-visual_2023} address temporal forgery localization, but they still focus on detecting manipulated segments rather than understanding the semantic inconsistencies across modalities.

More closely related to our work, AVFF~\cite{oorloff_avff_nodate} and AVoiD-DF~\cite{yang_avoid-df_2023} employ cross-modal classifiers for inconsistency detection. However, they target \textit{artifact-level} inconsistencies (e.g., lip-sync artifacts, facial blending defects) that can be identified via uni-modal artifacts. Our benchmark focuses on \textit{semantic-level} inconsistencies requiring high-level reasoning: e.g., a "male voice in a video showing a female speaker" violates identity, or "cheerful music in a funeral scene" violates emotion. These demand genuine cross-modal understanding. Additionally, existing deepfake datasets contain millisecond-level manipulations within short clips, insufficient for segment-level consistency evaluation. AVID addresses these gaps with 11.2K videos (avg. 235.5s) and 39.4K annotated segments (avg. 14.8s).

\section{The \modelname Benchmark}\label{sec:avbench}

To construct \modelname, we propose an efficient and scalable pipeline that consists of: (i) high-quality multimodal full-video sourcing, taxonomy definition, and segmentation (\S\ref{sec:data_source}--\S\ref{sec:temporal_seg}); (ii) agent-based strategy planning and injection-target selection (\S\ref{sec:strategy_agent}); and (iii) strategy-conditioned, multi-category inconsistency injection (\S\ref{sec:injectors}). The overall construction and annotation workflow is illustrated in Figure~\ref{fig:benchmark_pipeline2}. Additional implementation details are deferred to Appendix~\ref{sec:appendix_injectors} and Appendix~\ref{sec:appendix_prompts}.

\subsection{Data Sourcing and Inconsistency Definition}\label{sec:data_source}\label{sec:taxonomy}

We build \modelname on top of videos sourced from the LongVALE collection, with raw media originally derived from YouTube. For this work, we directly use their released original-video lists as our source pool. These videos have been curated by LongVALE to ensure semantic consistency between audio and visual content, and are verified for audio-visual alignment.

On top of these semantically consistent full videos, we formalize segment typing based on fundamental audio-visual factors. In the audio stream, we distinguish between \textbf{human speech} and \textbf{background sound} (ambient or background music, BGM). In the video stream, we distinguish between the \textbf{active speaker} and the \textbf{scene environment}. Based on these factors and their relationships, each segment is categorized into one of three semantic classes: \textbf{Class~1 (Active Speaker)}, where human speech is present and the speaking subject appears in the frame; \textbf{Class~2 (Voiceover)}, where human speech is present but the speaking subject is not visible; and \textbf{Class~3 (Scenic)}, where no human speech is present, and only background sound/BGM/silence is heard. The visual presence of a potential speaker does not affect the classification in this case.

Segment typing is crucial as inconsistency patterns are class-dependent. For example, Class~1 commonly involves issues like lip-sync mismatch, temporal offset, and speaker identity inconsistency; Class~2 focuses on semantic contradictions between narration and scene content, where strict lip-sync inconsistencies are less relevant; Class~3 highlights conflicts related to environmental and physical-acoustic factors. Based on these class-specific observations, we define a taxonomy of 8 inconsistency categories for construction and evaluation, which are detailed in Appendix Table~\ref{tab:class_mapping}.

This decouples \emph{where the sample comes from} (segment type) and \emph{how it is generated} from \emph{which relation is violated}, ensuring that the evaluation measures cross-modal reasoning rather than merely reverse-engineering synthesis artifacts.
\subsection{Temporal Segmentation}\label{sec:temporal_seg}

This stage establishes the temporal foundation for the benchmark. The input is a full-length raw video, and the goal is to segment it into semantically coherent, class-aware temporal units, each representing a specific segment type. These segments are then used reliably for downstream strategy planning and inconsistency injection.

The motivation for temporal segmentation is twofold. First, a full video typically contains multiple semantic events, and manipulation should only be applied to temporally valid regions. Second, different inconsistency categories depend on the segment class, so segment typing must occur before strategy selection. As shown in Figure~\ref{fig:benchmark_pipeline2}, we segment and classify the video into three distinct classes using a four-step pipeline:

\paragraph{Step 1: Speech Activity Detection (VAD).}  
We begin by extracting the audio stream and applying the Silero VAD model to detect voice activity and identify speech timestamps. This step creates the initial partition between speech and non-speech regions, providing the primary input for class-aware segmentation.

\paragraph{Step 2: Active-Speaker Proxy Detection (ASD).}  
Next, we sample video frames and apply the MediaPipe model for face-based active-speaker proxy detection. This step identifies periods where a visible speaking subject is present in the frame and distinguishes on-screen speaking content from off-screen narration or voiceover content.

\paragraph{Step 3: Timeline Construction and Class Assignment.}  
We combine the temporal boundaries from VAD and ASD outputs, sort them into a unified timeline, and classify each micro-segment based on two binary attributes: whether speech is present and whether a visible speaker is detected in the video. The resulting rule-based mapping assigns each segment to one of the following three semantic classes:  
\textbf{Class~1 (Active Speaker)}: The audio contains human speech \emph{and} the visual stream has a detected on-screen speaker. These segments are suitable for evaluating lip-sync consistency, voice identity alignment, and speaker-related manipulations.  
\textbf{Class~2 (Voiceover)}: The audio contains human speech, but no on-screen speaker is detected (e.g., off-screen narration or dubbed commentary). These segments are suitable for evaluating the semantic alignment between spoken content and visual scenes.  
\textbf{Class~3 (Scenic)}: The audio contains no human speech (only background music, ambient sounds, or silence). These segments are suitable for evaluating environmental, atmospheric, and spatial acoustic consistency.

\paragraph{Step 4: Temporal Fusion and Smoothing.}  
Due to the fine-grained behavior of the detectors, the raw outputs may contain fragmented segments, such as brief pauses within continuous speech being incorrectly classified as \textbf{Class~3 (Scenic)}. We therefore apply temporal fusion and smoothing by merging adjacent segments with identical labels, absorbing short non-speech fragments under duration constraints, and rounding timestamps for stable downstream processing. This produces cleaner and more reliable segment boundaries for subsequent strategy planning.

The final output of this stage is a timeline recording the segment boundaries, class labels, and confidence scores for each segment. Detailed class criteria and per-class statistics are summarized in Table~\ref{tab:seg_decision}.

\begin{table}[h]
\centering
\caption{Decision logic and per-class segment statistics for temporal segment classification.}
\label{tab:seg_decision}
\resizebox{\linewidth}{!}{
\begin{tabular}{ccccc}
\toprule
\textbf{Speech?} & \textbf{Active Face?} & \textbf{Class} & \textbf{\#Segments} & \textbf{Share (\%)} \\
\midrule
Yes & Yes & Class~1 (Active Speaker) & 32{,}908 & 41.80 \\
Yes & No  & Class~2 (Voiceover) & 18{,}096 & 22.99 \\
No  & N/A & Class~3 (Scenic) & 27{,}712 & 35.20 \\
\bottomrule
\end{tabular}
}
\end{table}


\subsection{Strategy Planning}\label{sec:strategy_agent}

After temporal segmentation, each full video is divided into class-aware segments. The next step is not to directly apply inconsistency injection, but to first determine \emph{whether} a segment should be manipulated and, if so, \emph{how} the inconsistency should be constructed. This step is critical because inconsistency construction depends on segment type, local semantics, and global video context. For example, lip-sync and voice-identity manipulations are only meaningful for \textbf{Active Speaker} segments, while semantic contradictions and background-sound conflicts are more appropriate for \textbf{Voiceover} and \textbf{Scenic} segments. Without such planning, applying injectors indiscriminately could lead to unnatural or weak inconsistencies.

To address this, we introduce a \textbf{Strategy Planning Agent}, powered by Gemini 3.1 Pro, which handles context-aware construction planning before inconsistency injection. As shown in Figure~\ref{fig:benchmark_pipeline2}, the agent considers three sources of information for each candidate segment: the segment class from temporal segmentation, the local segment content, and the defined inconsistency categories. Based on this information, the agent makes two sequential decisions. First, it performs \textbf{feasibility screening} to determine whether the segment is suitable for inconsistency construction, filtering out segments with poor content quality or weak semantic relevance. Second, for segments deemed feasible, the agent selects the most appropriate inconsistency category from the available options. The aim is to ensure that the injected inconsistency is both semantically plausible and compatible with the segment’s category, rather than a random perturbation.

This planning stage is essential as it bridges the class-aware temporal segmentation with the downstream injectors. It ensures that inconsistency construction is guided by semantic reasoning, rather than heuristic template matching alone. This results in a more diverse and realistic benchmark, with segment classes mapped to contextually appropriate inconsistency patterns through a unified decision process.

The strategy prompt is developed iteratively through a human-in-the-loop process, not a one-shot design. Annotators periodically review sampled strategies, identify failure cases, and provide feedback, which is used to refine the strategy prompt across multiple rounds. The final prompt is included in the appendix (Figure~\ref{fig:prompt_strategy_planning}). Additionally, all generated strategies undergo further human review before the actual injection. Details of this manual review process are provided in \S\ref{sec:human_effort}.

\begin{figure*}[t]
    \centering
    \begin{subfigure}[t]{0.36\textwidth}
        \centering
        \IfFileExists{static/number.pdf}{\includegraphics[width=\linewidth]{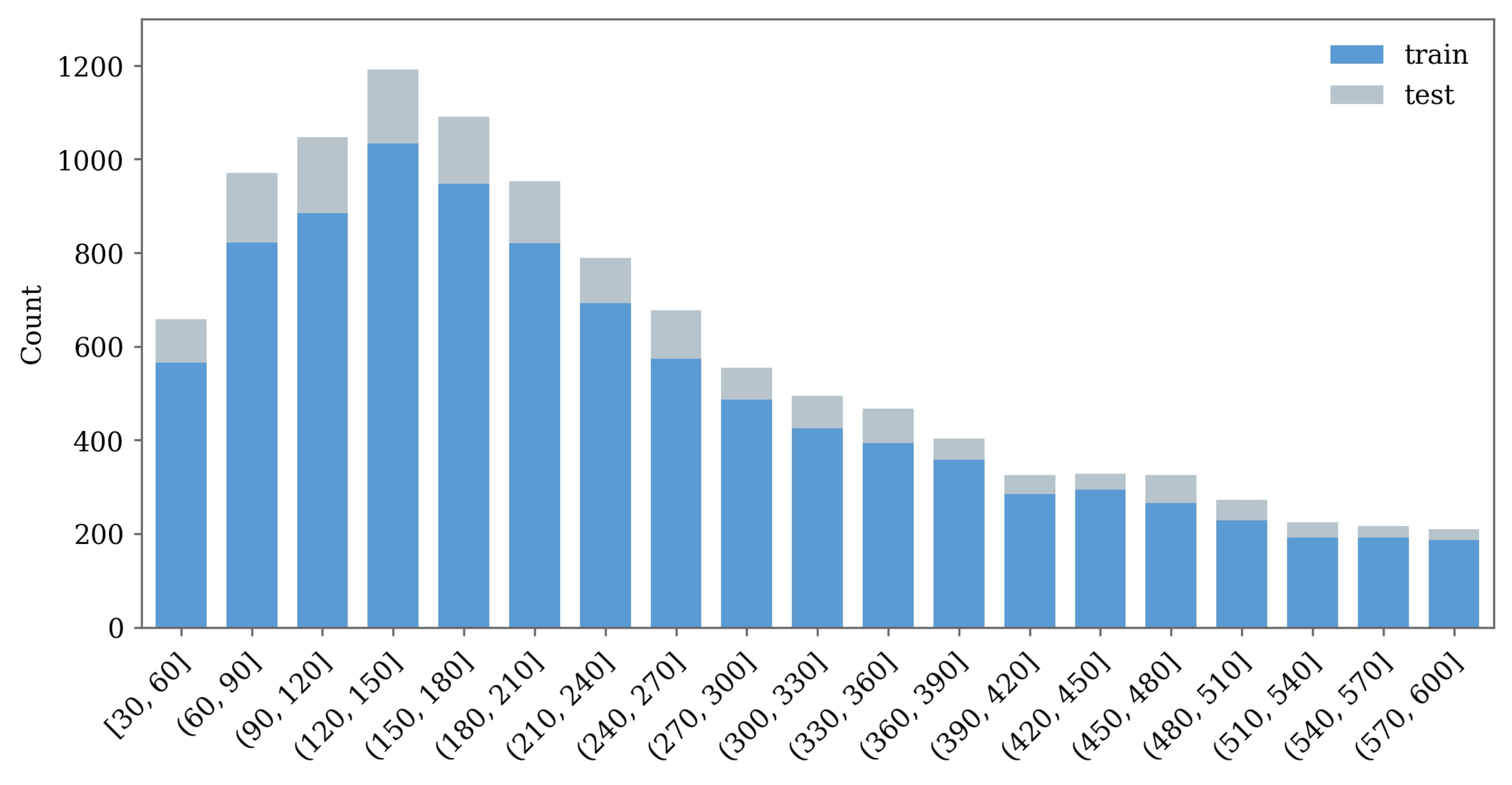}}{\fbox{\parbox[c][0.22\textheight][c]{0.95\linewidth}{\centering TODO: number.pdf}}}
        \caption{Video duration distribution (train/test)}
        \label{fig:stats_combined_a}
    \end{subfigure}
    \hfill
    \begin{subfigure}[t]{0.36\textwidth}
        \centering
        \IfFileExists{static/segment_duration.pdf}{\includegraphics[width=\linewidth]{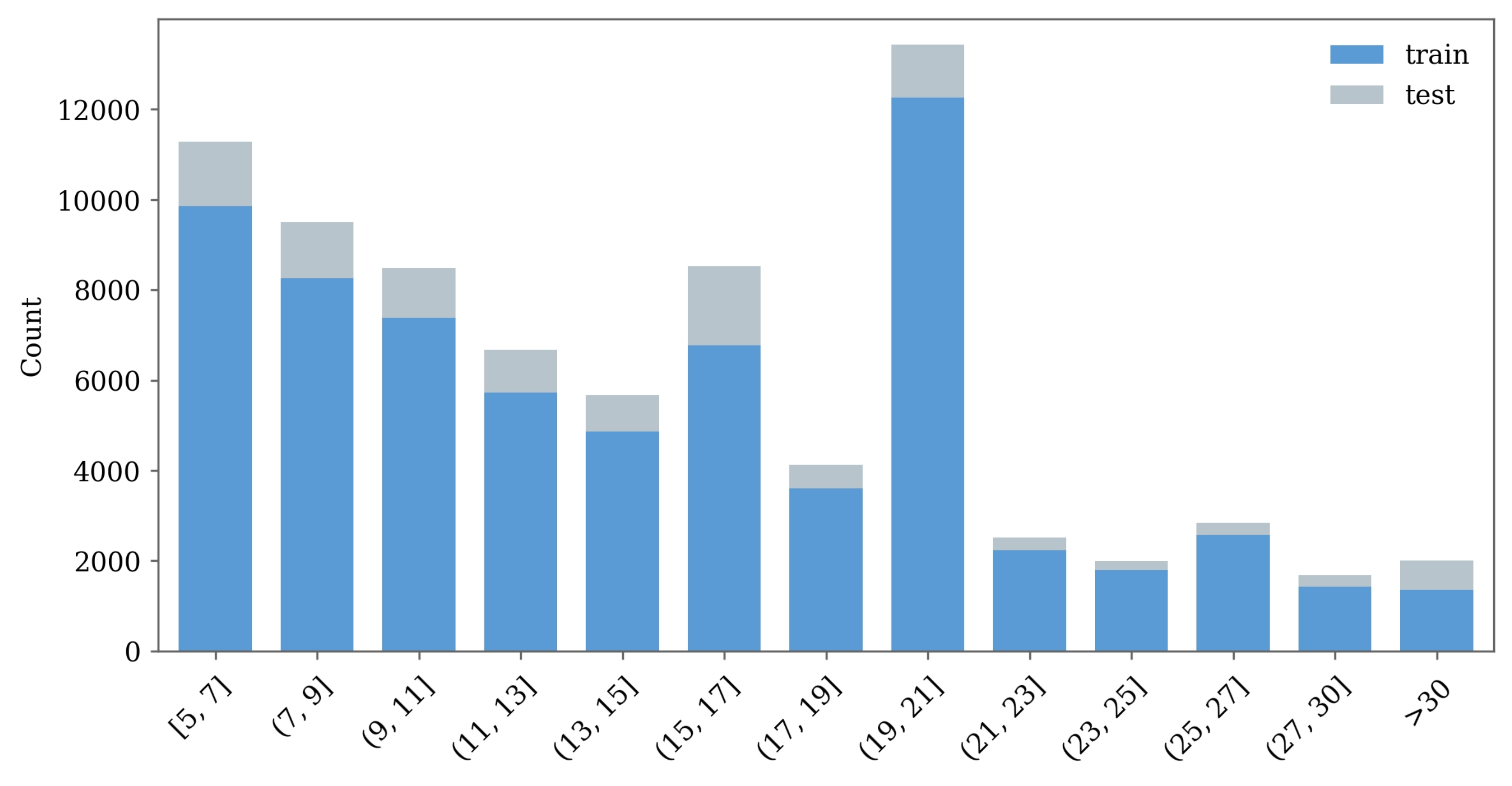}}{\fbox{\parbox[c][0.22\textheight][c]{0.95\linewidth}{\centering TODO: segment_duration.pdf}}}
        \caption{Segment duration distribution (train/test)}
        \label{fig:stats_combined_b}
    \end{subfigure}
    \hfill
    \begin{subfigure}[t]{0.24\textwidth}
        \centering
        \IfFileExists{static/event.pdf}{\includegraphics[width=0.9\linewidth]{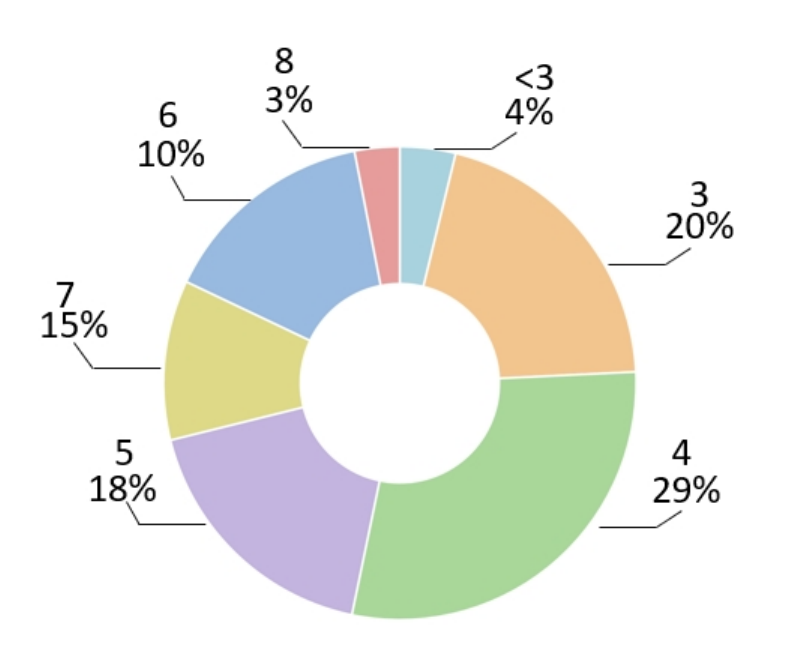}}{\fbox{\parbox[c][0.22\textheight][c]{0.95\linewidth}{\centering TODO: 事件数量分布.pdf}}}
        \caption{inconsistency events per video}
        \label{fig:stats_combined_c}
    \end{subfigure}

    \vspace{0.8em}

\begin{subfigure}[b]{0.24\textwidth}
    \raggedright
    \includegraphics[width=0.95\linewidth,height=0.18\textheight,keepaspectratio]{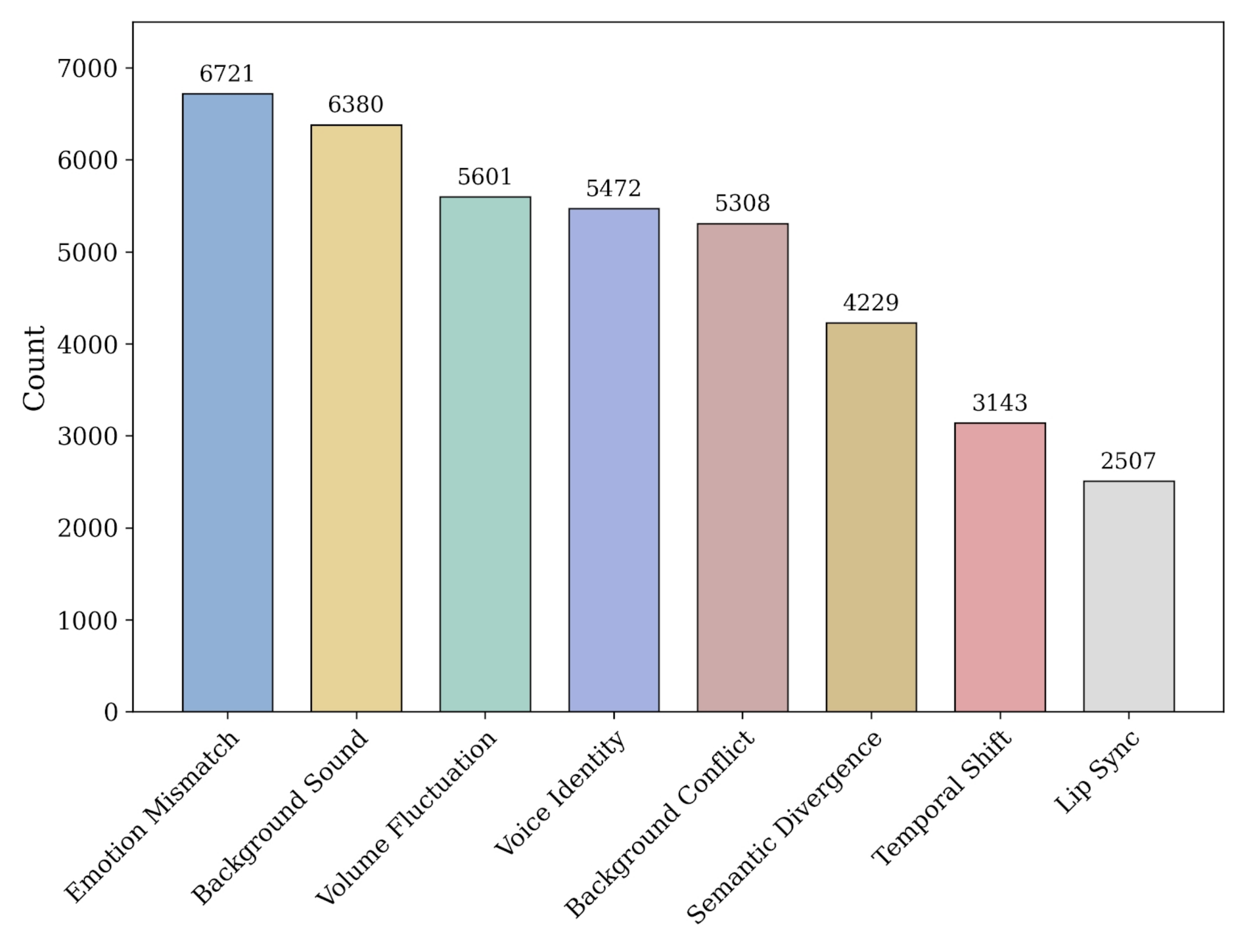}
    \caption{inconsistency categories}
    \label{fig:stats_combined_d}
\end{subfigure}
\hfill
\begin{subfigure}[b]{0.75\textwidth}
    \centering
\includegraphics[width=\linewidth,height=0.3\textheight,keepaspectratio]{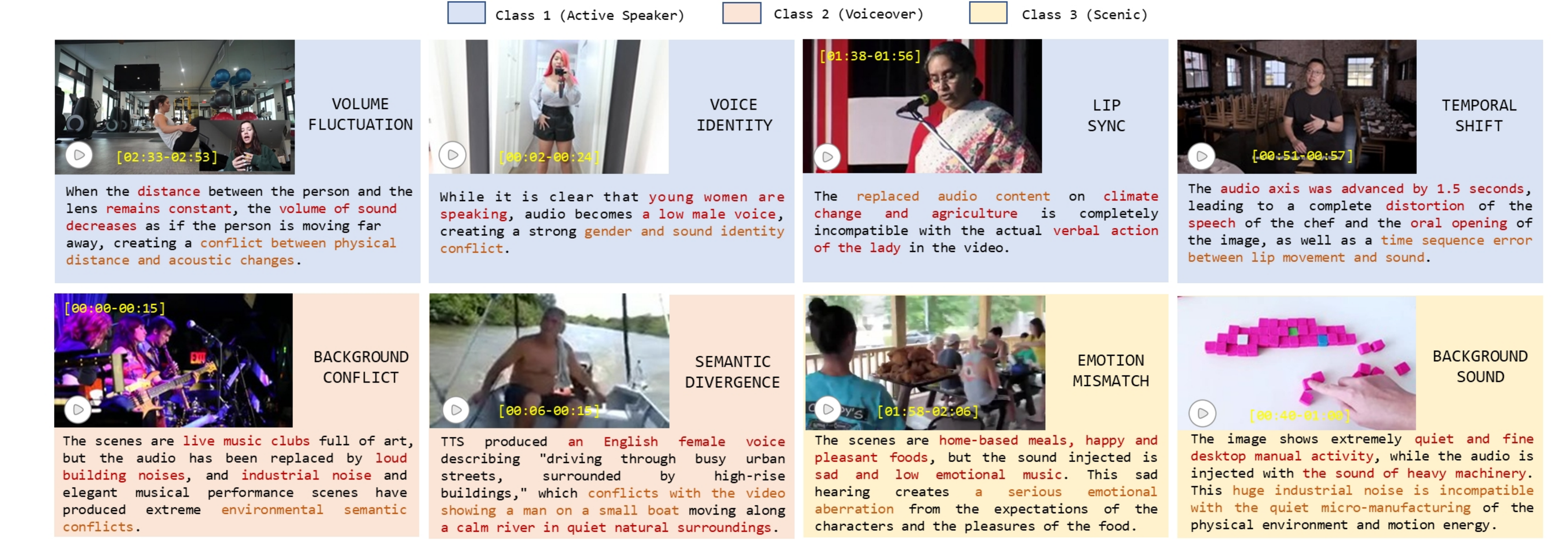}%
    \label{fig:stats_combined_e}
\end{subfigure}

    \caption{Comprehensive dataset static analysis: (a) Video duration distribution of both training and test sets, (b) segment duration distribution of both training and test sets, (c) pie chart of inconsistency events per video, (d) counts of 8 audio-visual inconsistency categories, and (e) examples of 8 audio-visual inconsistency categories.}
    \label{fig:stats_combined}
\end{figure*}

\subsection{Inconsistency Injection}\label{sec:injectors}

Once a strategy is selected in \S\ref{sec:strategy_agent}, the Execution Agent routes it to one of five specialized injectors. The agent generates the necessary executable parameters (e.g., offset magnitude, contradictory text, target voice profile, attenuation direction, background type) based on the plan, which are then passed to the corresponding injector to perform the inconsistency construction. Detailed implementation of each injector can be found in Appendix Table~\ref{tab:injectors}, and the mapping between inconsistency categories and injectors is summarized in Table~\ref{tab:class_mapping}.

The five injectors are designed to introduce different types of inconsistencies into the video:
\textbf{Temporal Injector} applies a temporal offset to the audio segment, adjusting synchronization between the audio and visual streams while preserving the original duration.  
\textbf{Semantic Injector} generates speech from contradictory text and synthesizes it to match the desired voice profile, mixing it with the original background to create semantic inconsistency.  
\textbf{Identity Injector} modifies the voice signal using pitch shifting, formant adjustments, and other techniques to simulate changes in age, gender, or identity, while maintaining perceptual consistency.  
\textbf{Spatial Injector} simulates spatial inconsistencies by applying volume attenuation to represent distance changes of the sound source, while the visual context remains unchanged.  
\textbf{Background Injector} replaces the environmental background with contradictory ambience or affective music, preserving the foreground speech if available, and generating inconsistencies related to the environment or emotional context. Specific details about the audio libraries used for this injector are provided in Appendix~\ref{sec:injector_details}.

All injectors operate under constraints to preserve the original video frames, resolution, and total duration. Only the audio within the selected windows is manipulated. Specific details about the implementation and tools used for each injector are provided in Appendix~\ref{sec:injector_details}.

\subsection{Human Effort and Subset Split}\label{sec:human_effort}

Human effort is integrated at two stages, contributing to the reliability of \modelname.

\paragraph{Stage I: Strategy-level Human Feedback and Filtering.}  
At the strategy stage, annotators review sampled strategy outputs, including feasibility decisions, injection choices, and content-level rationales. They flag invalid or low-plausibility plans, provide corrective feedback, and update instruction constraints. These feedback signals are incorporated iteratively into prompt updates across multiple rounds of LLM refinement (final prompt provided in Appendix Figure~\ref{fig:prompt_strategy_planning}). Specifically, after every 50 strategies are generated, audio-visual experts review them, flagging unreasonable injection strategies (e.g., when the injected inconsistency is not easily identifiable by humans as inconsistent). The strategy balance is also monitored to prevent an over-representation of any single category. The process continues until fewer than 3 unreasonable strategies are identified in each set of 50, achieving a 95\% accuracy rate. Once the prompt is finalized, it is used for the strategy agent. Generated strategies undergo manual filtering before execution, with only approved strategies dispatched to injectors.

\paragraph{Stage II: Post-construction Video-level Quality Control.}
After automatic sanity checks, we manually inspect the constructed videos to remove weak or ambiguous samples and correct annotation errors. The core acceptance principle is human verifiability: annotators should be able to determine if a video is consistent or inconsistent, identify the inconsistency category if applicable, and verify whether the assigned label matches the observed evidence. Videos with poor construction quality or substantial deviations from human interpretation are filtered out.

To quantify annotation reliability, we conducted inter-annotator agreement studies. As detailed in Appendix~\ref{sec:appendix_annotation_quality}, three annotators independently labeled each sample, achieving substantial agreement (Cohen's Kappa > 0.75 across all tasks). These results confirm the reliability of AVID annotations.

Through this scalable pipeline, we construct a dataset of 8,400 full videos with inconsistent events, generating 39,361 inconsistent video segments. The full video dataset consists of 11,200 videos (3:1 ratio of inconsistent to consistent samples), split into 9,639/1,561 for training and testing, ensuring consistency in distributions such as video length and event count. For the segment dataset, we maintain a 1:1 ratio between inconsistent and consistent segments, totaling 78,722 segment videos. This dataset is split into 68,088/10,634 for training and testing, ensuring consistent distribution.

\subsection{Statistical Analysis}\label{sec:stats}

We provide a statistical overview of \modelname{} and highlight its key characteristics, as shown in Fig.~\ref{fig:stats_combined}. Overall, \modelname{} contains 11,200 videos with 39,361 annotated inconsistency events, making it a large-scale benchmark with dense temporal annotations and diverse inconsistency patterns. \modelname{} supports both segment-level and full-video evaluation, enabling the assessment of localized inconsistency recognition as well as long-range temporal reasoning.

\paragraph{Segment-level Dataset.}
The segment-level dataset consists of inconsistent and corresponding consistent segments, extracted from longer videos. These segments generally range from 5 to 20 seconds, with some exceeding 20 seconds. As shown in Fig.~\ref{fig:stats_combined_b}, the distribution of segment durations is balanced. These segments serve as the minimal units for detecting inconsistencies inherent to the video content. Captions for each segment indicate whether the segment is inconsistent and provide detailed reasoning. The distribution of inconsistency categories is balanced, with specific categories shown in Fig.~\ref{fig:stats_combined_d}. The segment-level dataset emphasizes detecting cross-modal inconsistencies based on self-contained evidence, facilitating fine-grained inconsistency recognition.

\paragraph{Full-video Dataset.}
The full-video dataset consists of long-form videos that contain multiple inconsistency events, alongside completely consistent segments. These videos range in duration from 30 to 600 seconds, as shown in Fig.~\ref{fig:stats_combined_a}. Most videos fall within the 1–3 minute range, while a significant portion extends beyond 3 minutes, providing ample temporal context for comprehensive inconsistency analysis. Inconsistency events are densely distributed, with \textbf{78\%} of videos containing 3 to 6 events, and only \textbf{4\%} containing fewer than 3 events (Fig.~\ref{fig:stats_combined_c}). Additionally, \textbf{18\%} of videos feature more than 7 inconsistencies, ensuring a mix of moderate and challenging examples. Each video is annotated for inconsistencies (yes/no), with timestamps and detailed reasoning for each event. Example annotations are shown in Fig.~\ref{fig:stats_combined}. Unlike the segment-level dataset, which only requires inconsistency detection at the segment level, the full-video dataset involves identifying multiple inconsistent segments within a full video and providing reasons for each inconsistency. This dense reasoning task is more challenging and requires the ability to perform fine-grained inconsistency recognition at the segment level.

\begin{figure}[t]
    \centering
    \includegraphics[width=1\linewidth]{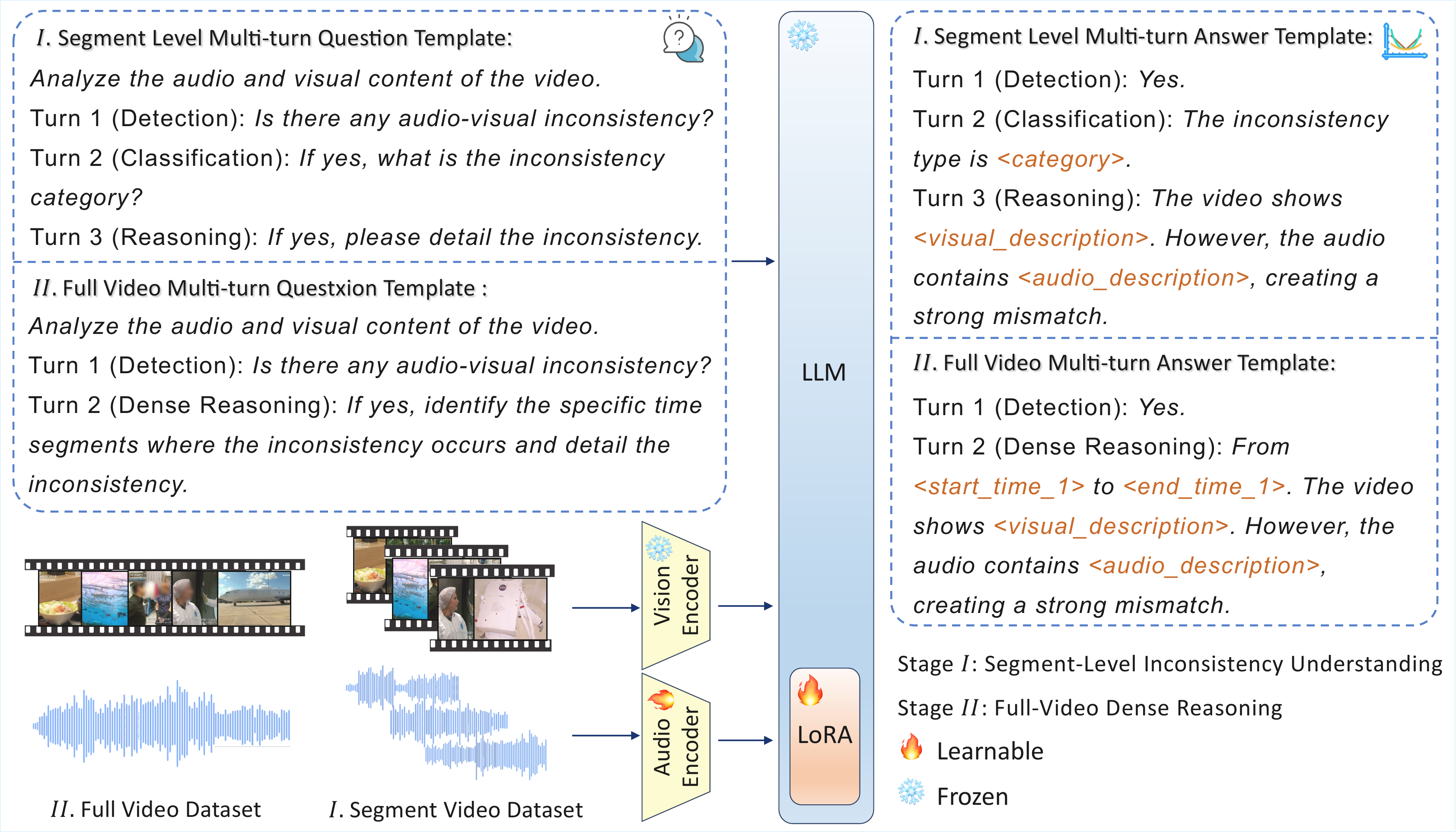}
    \caption{The \baselinename architecture for omni-modal audio-visual inconsistency detection and reasoning.}
    \label{fig:avbench_architecture}
\end{figure}

\begin{table*}[htbp]
    \centering
    \small
    \setlength{\tabcolsep}{2.5pt} 
    \newcommand{\cw}[1]{\makebox[0.7cm][c]{#1}}

    \caption{Comparison with existing models on the Segment and Full-Video tasks of the AVID benchmark. \textbf{Detect}: identify audio-visual inconsistency. \textbf{Cla.}: classify inconsistency category. \textbf{R@$\alpha$}: Recall@1 at IoU threshold $\alpha$. \textbf{mIoU}: mean Intersection over Union. \textbf{B/R/M}: BLEU-4, ROUGE-L, and METEOR. \textbf{$S_M$}: SODA-m score reflecting holistic dense understanding. \textbf{All metrics are reported in percentage scale (values $\times 100$).}}
    \label{tab:model_comparison}
    \begin{tabular*}{\textwidth}{@{\extracolsep{\fill}} l | cc ccc | c cccc ccc c @{}}
        \toprule
        \multirow{3}{*}{\textbf{Model}} & \multicolumn{5}{c|}{\textbf{Segment}} & \multicolumn{9}{c}{\textbf{Full-Video}} \\
        \cmidrule{2-15}
        & \cw{\textbf{Detect}} & \cw{\textbf{Cla.}} & \multicolumn{3}{c|}{\textbf{Reasoning}} & \cw{\textbf{Detect}} & \multicolumn{4}{c}{\textbf{Grounding}} & \multicolumn{3}{c}{\textbf{Reasoning}} & \cw{\textbf{Dense}} \\
        \cmidrule{2-15}
        & \cw{Acc} & \cw{Acc} & \cw{B} & \cw{R} & \cw{M} & \cw{Acc} & \cw{R@0.3} & \cw{R@0.5} & \cw{R@0.7} & \cw{mIoU} & \cw{B} & \cw{R} & \cw{M} & \cw{$S_M$} \\
        \midrule
        Arc-Hunyuan-Video & 41.6 & 11.1 & 1.4 & 16.3 & 17.0 & 52.3 & 9.8  & 6.2  & 2.8  & 6.4  & 0.01 & 1.55 & 0.73 & 0.57 \\
        OLA                 & 48.1 & 7.9  & 1.3 & 15.4 & 11.9 & 59.3 & 14.3 & 7.1  & 2.8  & 8.8  & 0.17 & 3.64 & 1.94 & 1.20 \\
        SALMONN 2 -7B     & 49.9   & 50.0 & 1.5  & 20.9   & 20.8 & 25.1 & 25.0 & 25.0 & 0.0  & 13.8 & 0.16 & 6.45 & 4.95 & 0.38 \\
        SALMONN 2 -72B     & 51.5   & 32.3   & 1.62  & 16.5   & 17.1   & 25.0   & -   & -   & -   & -   & -   & -   & -   & -   \\
        Qwen3-Omni          & 52.6 & 55.0 & 2.2 & 21.7 & 23.8 & 75.3 & 12.3 & 8.3  & 4.9  & 9.1  & 0.25 & 3.90 & 2.48 & 1.48 \\
        MiMo-V2-Omni        & 60.5 & 53.6 & 2.6 & 22.6 & 22.8 & 63.9 & 23.3  & 22.2  & 18.9  & 19.5  & 0.53 & 5.50 & 5.08 & 5.37 \\
        Gemini 2.5 Pro    & 68.2 & 56.4 & 2.2 & 20.6 & 22.1 & 81.6 & 28.0 & 25.9 & 23.8 & 24.3 & 0.58 & 6.46 & 5.12 & 5.83 \\
        Gemini 3.1 Pro    & 69.7 & 57.1 & 2.5 & 21.2 & 22.6 & 84.9 & 30.9 & 28.1 & 25.4 & 26.2 & 0.64 & 7.04 & 5.49 & 6.15 \\

        \midrule

        \textbf{AVID-Qwen} & \textbf{61.3} & \textbf{55.5} & \textbf{6.2} & \textbf{26.5} & \textbf{25.9} & \textbf{78.2} & \textbf{55.2} & \textbf{39.2} & \textbf{19.1} & \textbf{36.1} & \textbf{2.73} & \textbf{15.2} & \textbf{15.0} & \textbf{7.47} \\
        \bottomrule
    \end{tabular*}
\end{table*}

\section{\baselinename: A Strong Omni-Modal Baseline}\label{sec:avbench_llm}

To demonstrate the learnability of \modelname and provide a reproducible reference point, we build a strong baseline by fine-tuning an open-weights omni-modal model on our benchmark protocol. We denote this model as \baselinename.

\subsection{Backbone Model}\label{sec:avbench_arch}

We adopt \textbf{Qwen3-Omni-30B-A3B-Instruct} as the backbone model. Qwen3-Omni is an omni-modal large language model that natively supports interleaved video (visual frames) and audio streams as input, and generates autoregressive textual outputs. The model architecture consists of three core components: (i) a vision encoder that processes video frames into visual tokens, (ii) an audio encoder that extracts acoustic features from the audio stream, and (iii) an LLM backbone that performs multimodal reasoning over concatenated visual and audio tokens. Detailed architecture specifications, including input preprocessing configurations (e.g., frame sampling rate, resolution), are provided in Appendix~\ref{sec:appendix_model_arch}.

The model takes video frames and a textual instruction as input, and generates an autoregressive textual response. We design task-specific prompts for different evaluation scenarios: (i) segment-level inconsistency detection, which requires determining whether a short clip contains audio-visual inconsistency, identifying the inconsistency category, and providing reasoning; and (ii) full-video inconsistency detection, which requires determining whether a full video contains any inconsistency, localizing each inconsistency event with temporal boundaries, and providing reasoning.

\subsection{Two-Stage Fine-Tuning}\label{sec:avbench_sft}

Zero-shot omni-models often struggle with fine-grained temporal grounding in long-form videos, while direct full-video training may lead to suboptimal localization accuracy. To address this, we propose a \textbf{progressive two-stage fine-tuning strategy}.

The core insight is that detecting inconsistencies in full videos requires two fundamental capabilities: (i) the ability to determine whether a given segment contains audio-visual inconsistency and identify its type, and (ii) the ability to localize multiple inconsistency events within a full video and provide dense reasoning. These two capabilities are learned most effectively through progressive training.

In the first stage, we fine-tune the model on segment-level data. Each training sample is a short video clip (5--20 seconds) that may or may not contain inconsistency. The model learns to (i) determine whether the segment contains audio-visual inconsistency, (ii) identify the inconsistency category from our 8-category taxonomy, and (iii) provide reasoning for the judgment. This stage establishes the fundamental capability of detecting and understanding inconsistencies at the segment level, which serves as the prerequisite for localizing inconsistencies in longer videos.

In the second stage, we continue fine-tuning on full-video data. Each training sample is a full video (30--600 seconds) that may contain multiple inconsistency events. Leveraging the segment-level detection capability from Stage 1, the model learns to (i) identify whether a full video contains any inconsistency, (ii) localize each inconsistency event with precise temporal boundaries (start and end timestamps), and (iii) provide reasoning for each detected event. This stage enables the model to perform dense temporal grounding and reasoning across long-form content.

The two-stage approach first builds the foundational detection capability, then extends it to multi-event localization and reasoning in full videos. The complete training hyper-parameters are provided in Appendix~\ref{sec:appendix_train_hparams}.

The resulting \baselinename serves as a transparent reference for comparing both open-source and closed-source omni-models under the same benchmark protocol.


\section{Experiments}\label{sec:experiments}

\subsection{Experimental Setup}\label{sec:exp_setup}
This section outlines the evaluation settings, model comparisons, evaluation protocol, and metrics.

\noindent\textbf{Compared Models.} We evaluate both closed-source and open-source omni-models under a unified few-shot protocol with the same prompt template (see Appendix~\ref{sec:appendix_prompts_eval}). Models are categorized as follows:
- \textbf{Closed-source models}: Gemini 2.5 Pro, Gemini 3.1 Pro, and MiMo-V2-Omni (accessed via API).
- \textbf{Open-source models}: Qwen3-Omni, OLA, Video-SALMONN 2 (7B, 72B), and ARC-Hunyuan-Video (self-hosted).
- \textbf{Our method}: \textbf{AVID-Qwen}, fine-tuned on both segment-level and full-video data (Stage 1 + Stage 2) as detailed in Section~\ref{sec:avbench_llm}, based on Qwen3-Omni.
Model configurations, including API endpoints, versions, temperature settings, and token limits, are provided in Appendix~\ref{sec:appendix_llm_config}.

\noindent\textbf{Evaluation Protocol.} We evaluate all models on two tasks using AVID: \textbf{segment-level tasks} and \textbf{full-video tasks}. Segment-level tasks assess inconsistency detection, category classification, and reasoning, while full-video tasks focus on detection, temporal grounding, and reasoning. The task design follows a coarse-to-fine approach: models first detect inconsistencies, then classify their type or segment grounding, and finally provide reasoning. Therefore, classification, grounding, and reasoning metrics are computed only for correctly detected inconsistencies (True Positives).

\noindent\textbf{Metrics.} We report Accuracy for inconsistency detection, BLEU-4, ROUGE-L, and METEOR for reasoning quality, and Recall@1 at IoU thresholds \{0.3, 0.5, 0.7\}, mIoU, and SODA-m for full-video temporal grounding and dense understanding. Detailed metric definitions and computation protocols are provided in Appendix~\ref{sec:appendix_metrics}.

\subsection{Main Results}\label{sec:exp_main}

Table~\ref{tab:model_comparison} presents the comprehensive evaluation results of state-of-the-art omni-models on both segment-level and full-video tasks. We analyze the results from two perspectives: (1) overall trends across model categories, and (2) task-specific performance analysis.

\noindent\textbf{Overall Trends.}
Our evaluation reveals clear performance patterns across model categories. Closed-source models (Gemini 2.5 Pro, Gemini 3.1 Pro, MiMo-V2-Omni) generally lead in detection accuracy — Gemini 3.1 Pro achieves 69.7\% for segment and 84.9\% for full video, the highest among all models. Open-source models show more varied performance: Qwen3-Omni demonstrates balanced capabilities across tasks, while OLA and ARC-Hunyuan-Video struggle with fine-grained understanding. Video-SALMONN 2 exhibits a unique behavior of rarely predicting inconsistencies, resulting in misleadingly high classification metrics (50.0\% for 7B) that should be interpreted with caution (see Appendix~\ref{sec:appendix_salmonn_analysis} for detailed analysis).

Notably, AVID-Qwen (fine-tuned on AVID training data) achieves the best overall performance across all models. It attains 2.8$\times$ higher BLEU-4 (6.2 vs 2.2) in segment reasoning and 4.5$\times$ higher BLEU-4 (2.73 vs 0.25) in full-video reasoning compared to its base model Qwen3-Omni, while also leading in temporal grounding (mIoU: 36.1\% vs 26.2\% for Gemini 3.1 Pro) and holistic understanding (SODA-m: 7.47 vs 6.15). These results validate that our benchmark provides valuable training resources for enhancing fine-grained audio-visual inconsistency understanding.

\noindent\textbf{Segment-level Performance.}
We evaluate three tasks at the segment level: detection, category classification, and reasoning.
\textbf{Detection.} Gemini 3.1 Pro achieves 69.7\% accuracy, followed by Gemini 2.5 Pro (68.2\%) and MiMo-V2-Omni (60.5\%) — these results show that closed-source models maintain a clear advantage in detection. Among open-source models, Qwen3-Omni reaches 52.6\%, while OLA (48.1\%) and ARC-Hunyuan-Video (41.6\%) lag behind, indicating substantial room for improvement in open-source detection capabilities.
\textbf{Classification.} Gemini models achieve 56.4\%$\sim$57.1\% accuracy, demonstrating strong fine-grained categorization ability. Qwen3-Omni (55.0\%) and MiMo-V2-Omni (53.6\%) show competitive performance, comparable to closed-source models. In contrast, OLA (7.9\%) and ARC-Hunyuan-Video (11.1\%) struggle with fine-grained categorization, suggesting these models lack sufficient training data for inconsistency category discrimination.
\textbf{Reasoning.} Gemini 3.1 Pro achieves BLEU-4/ROUGE-L/METEOR of 2.5/21.2/22.6, while Qwen3-Omni performs comparably at 2.2/21.7/23.8 — indicating that open-source models can match closed-source reasoning on this task. However, OLA (1.3/15.4/11.9) and ARC-Hunyuan-Video (1.4/16.3/17.0) show significantly weaker reasoning capabilities, revealing a substantial gap among open-source models.

\noindent\textbf{Full-video Performance.}
The full-video task involves detection, temporal grounding, and reasoning on correctly grounded samples.
\textbf{Detection.} Gemini models achieve 81.6\%$\sim$84.9\% accuracy, significantly outperforming Qwen3-Omni (75.3\%) and MiMo-V2-Omni (63.9\%). Open-source models like OLA (59.3\%) and ARC-Hunyuan-Video (52.3\%) show even lower detection rates, highlighting the detection challenge in full-video scenarios.
\textbf{Temporal Grounding.} Gemini 3.1 Pro achieves mIoU of 26.2\% (R@0.5: 28.1\%) — the best grounding performance among all models. MiMo-V2-Omni follows at 19.5\% (R@0.5: 22.2\%). In contrast, Qwen3-Omni achieves only 9.1\% mIoU (R@0.5: 8.3\%) — a 2.9$\times$ gap compared to Gemini 3.1 Pro. OLA and ARC-Hunyuan-Video perform even worse, demonstrating that temporal grounding remains a critical challenge for current omni-models.
\textbf{Reasoning.} Gemini 3.1 Pro leads with BLEU-4/ROUGE-L/METEOR of 0.64/7.04/5.49 and SODA-m of 6.15. MiMo-V2-Omni follows at 0.53/5.50/5.08 and 5.37. Open-source models show weaker reasoning: Qwen3-Omni achieves only 0.25/3.90/2.48 — a 2.6$\times$ gap in BLEU-4 compared to Gemini 3.1 Pro. OLA and ARC-Hunyuan-Video perform even lower, confirming that reasoning on full videos is particularly challenging for current models.

These results demonstrate that current omni-models face challenges in fine-grained audio-visual inconsistency understanding, particularly in reasoning and temporal grounding tasks.

\begin{figure}[t]
    \centering
    \begin{minipage}[t]{0.24\linewidth}
        \centering
        \includegraphics[width=\linewidth]{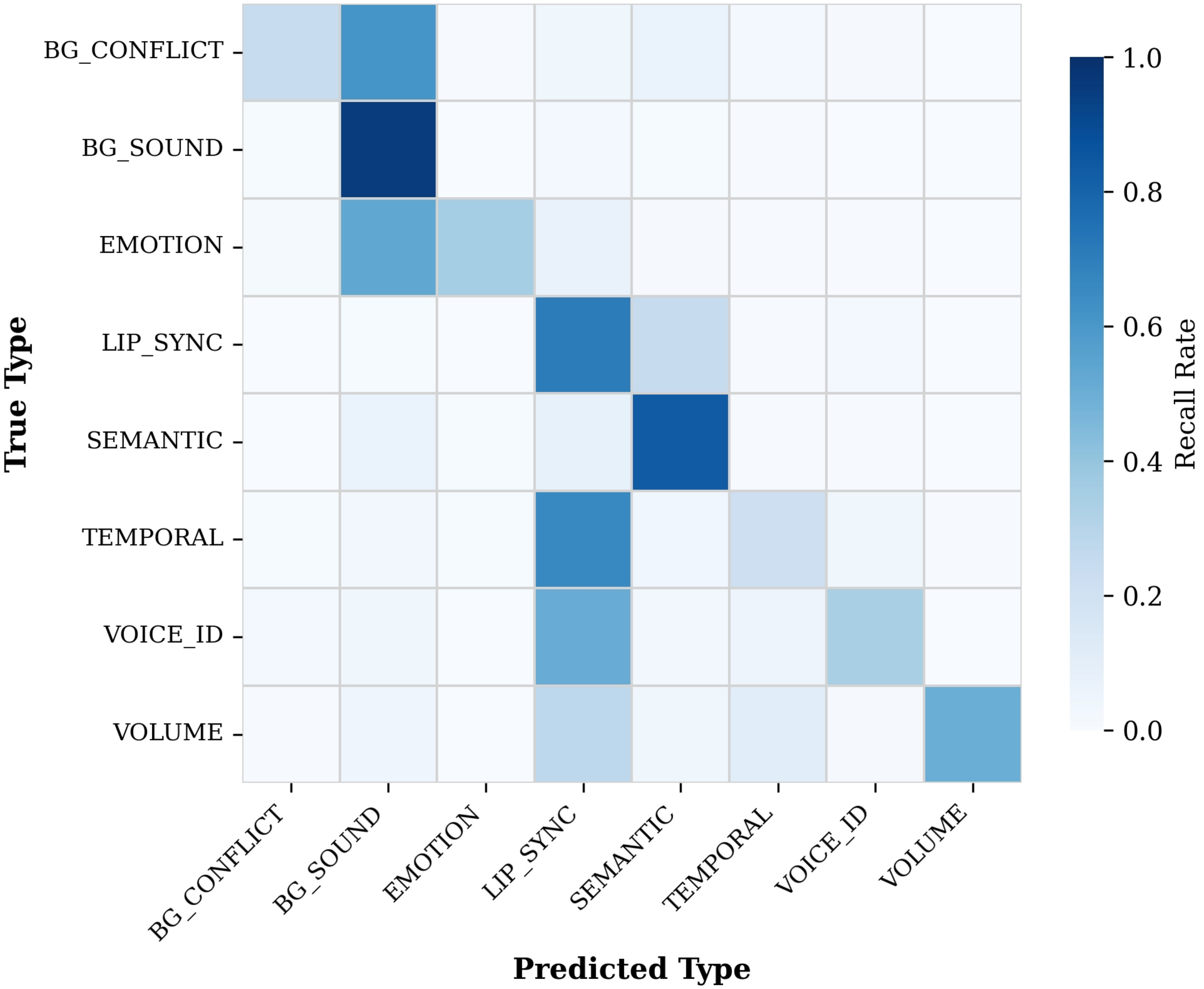}
        \\[-0.5em]
        {\small (a) Gemini 3.1 Pro}
    \end{minipage}\hfill
    \begin{minipage}[t]{0.24\linewidth}
        \centering
        \includegraphics[width=\linewidth]{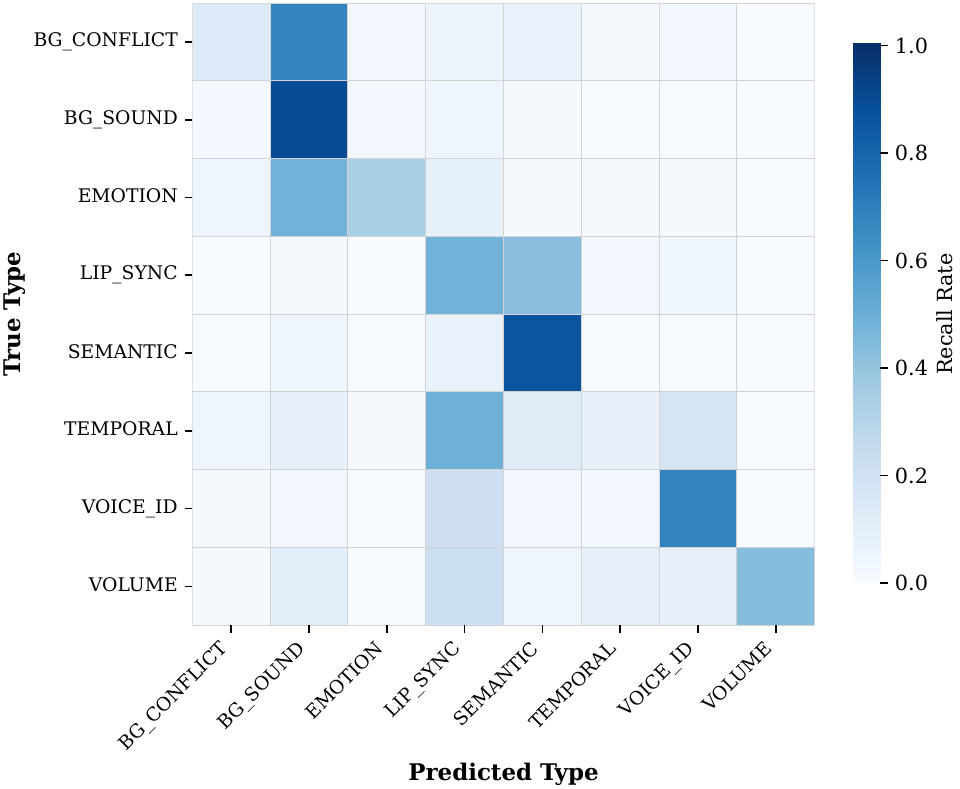}
        \\[-0.5em]
        {\small (b) Gemini 2.5 Pro}
    \end{minipage}\hfill
    \begin{minipage}[t]{0.24\linewidth}
        \centering
        \includegraphics[width=\linewidth]{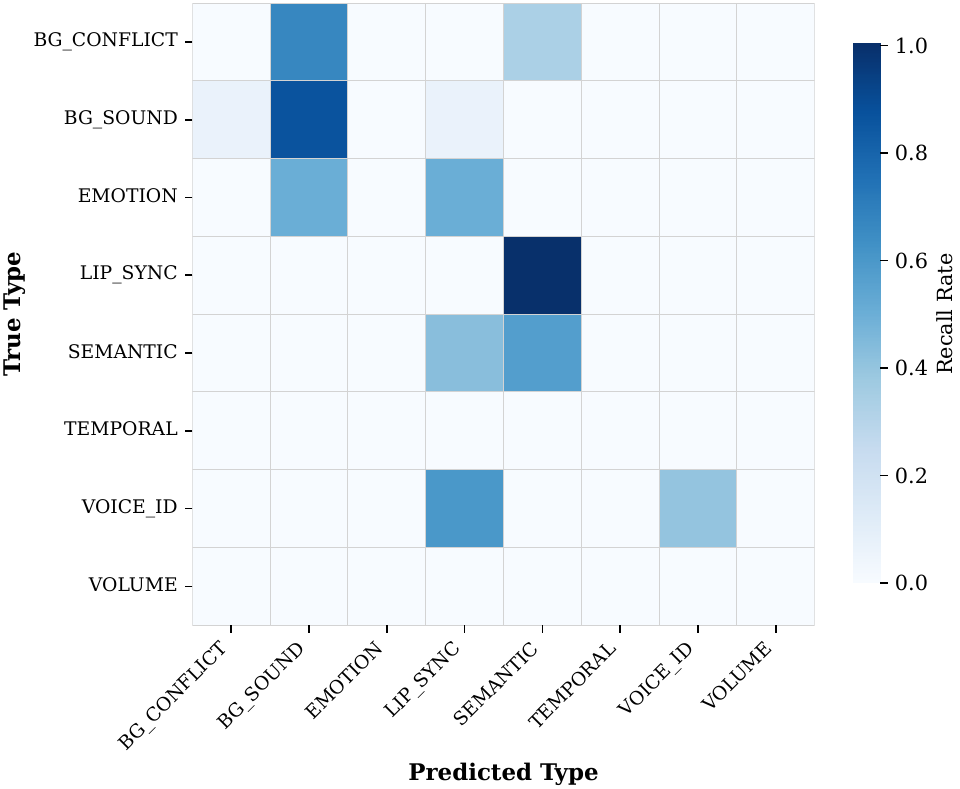}
        \\[-0.5em]
        {\small (c) MiMo-V2}
    \end{minipage}\hfill
    \begin{minipage}[t]{0.24\linewidth}
        \centering
        \includegraphics[width=\linewidth]{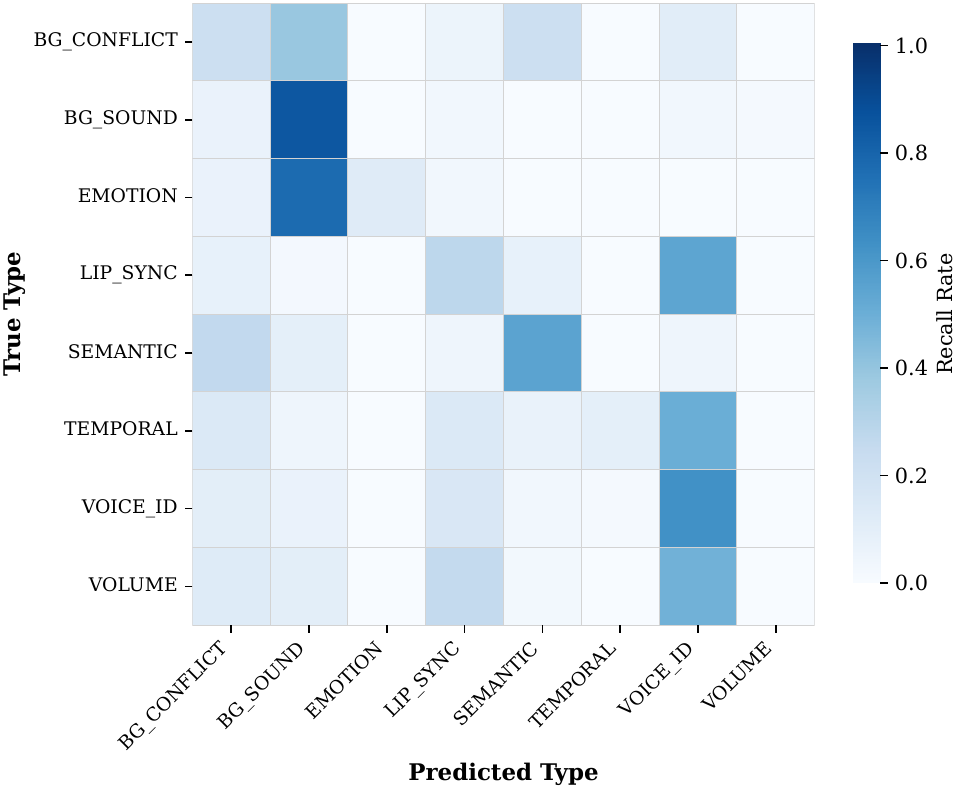}
        \\[-0.5em]
        {\small (d) AVID-Qwen}
    \end{minipage}

    \vspace{0.4em}

    \begin{minipage}[t]{0.24\linewidth}
        \centering
        \includegraphics[width=\linewidth]{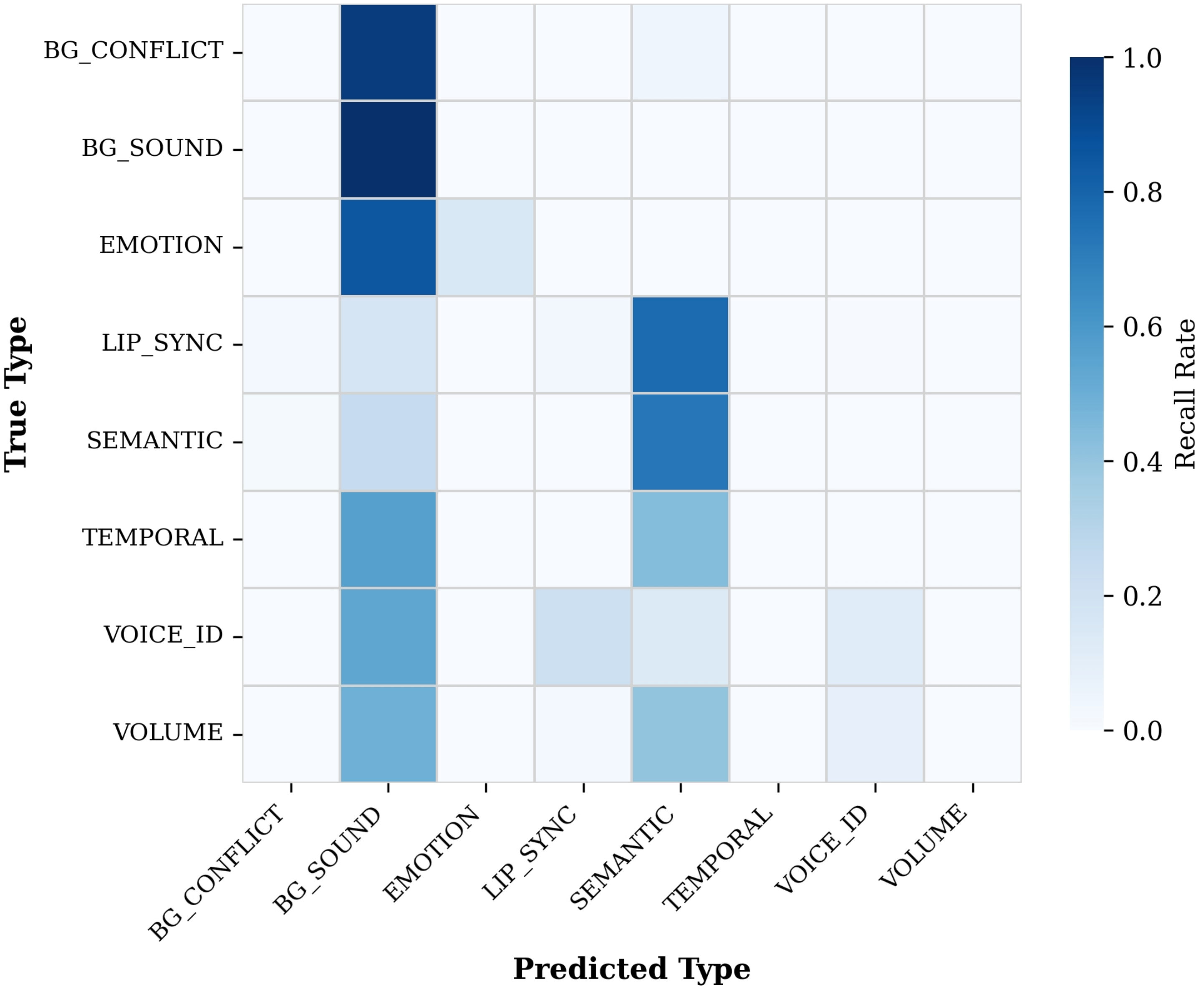}
        \\[-0.5em]
        {\small (e) Qwen3-Omni}
    \end{minipage}\hfill
    \begin{minipage}[t]{0.24\linewidth}
        \centering
        \includegraphics[width=\linewidth]{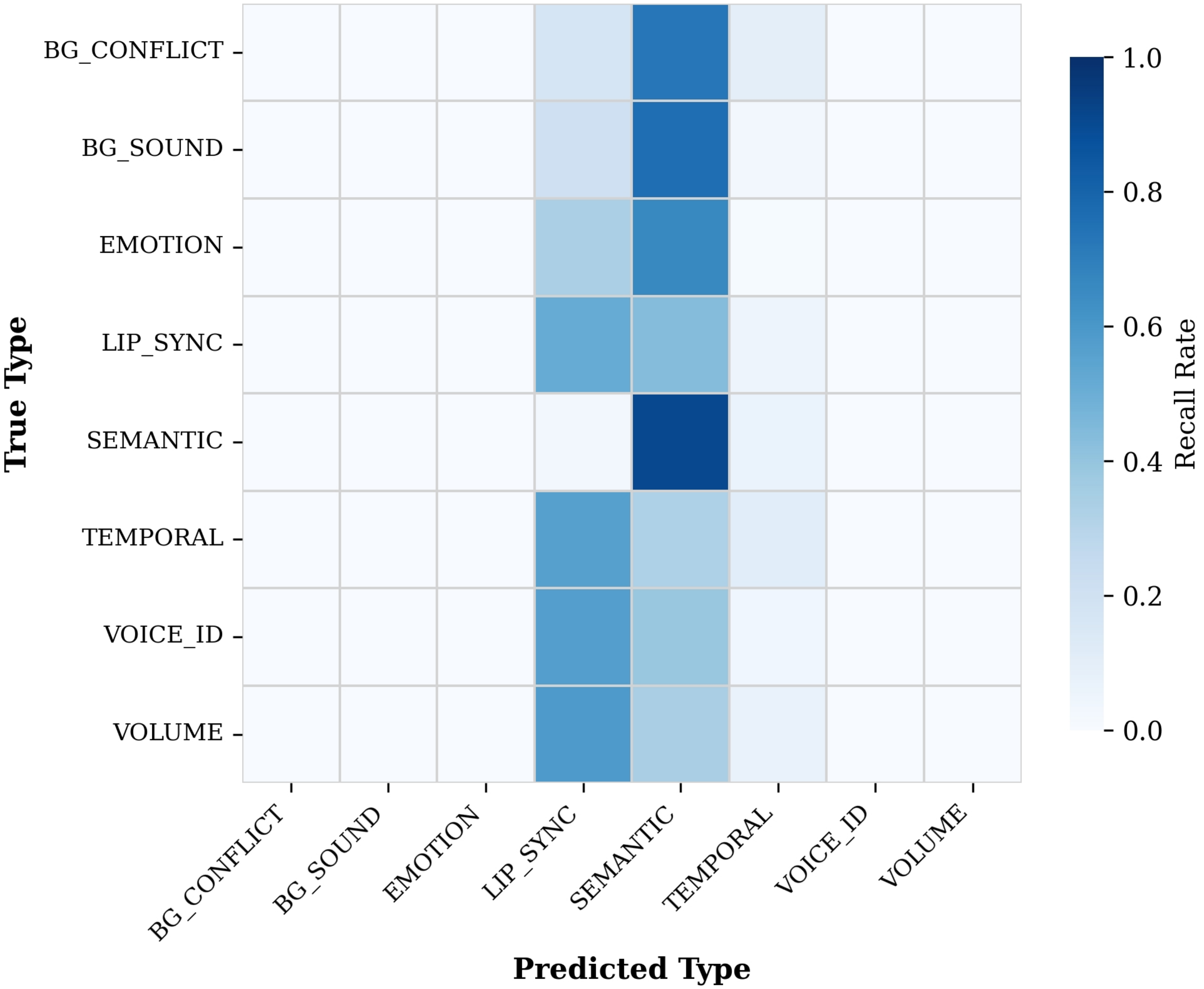}
        \\[-0.5em]
        {\small (f) OLA}
    \end{minipage}\hfill
    \begin{minipage}[t]{0.24\linewidth}
        \centering
        \includegraphics[width=\linewidth]{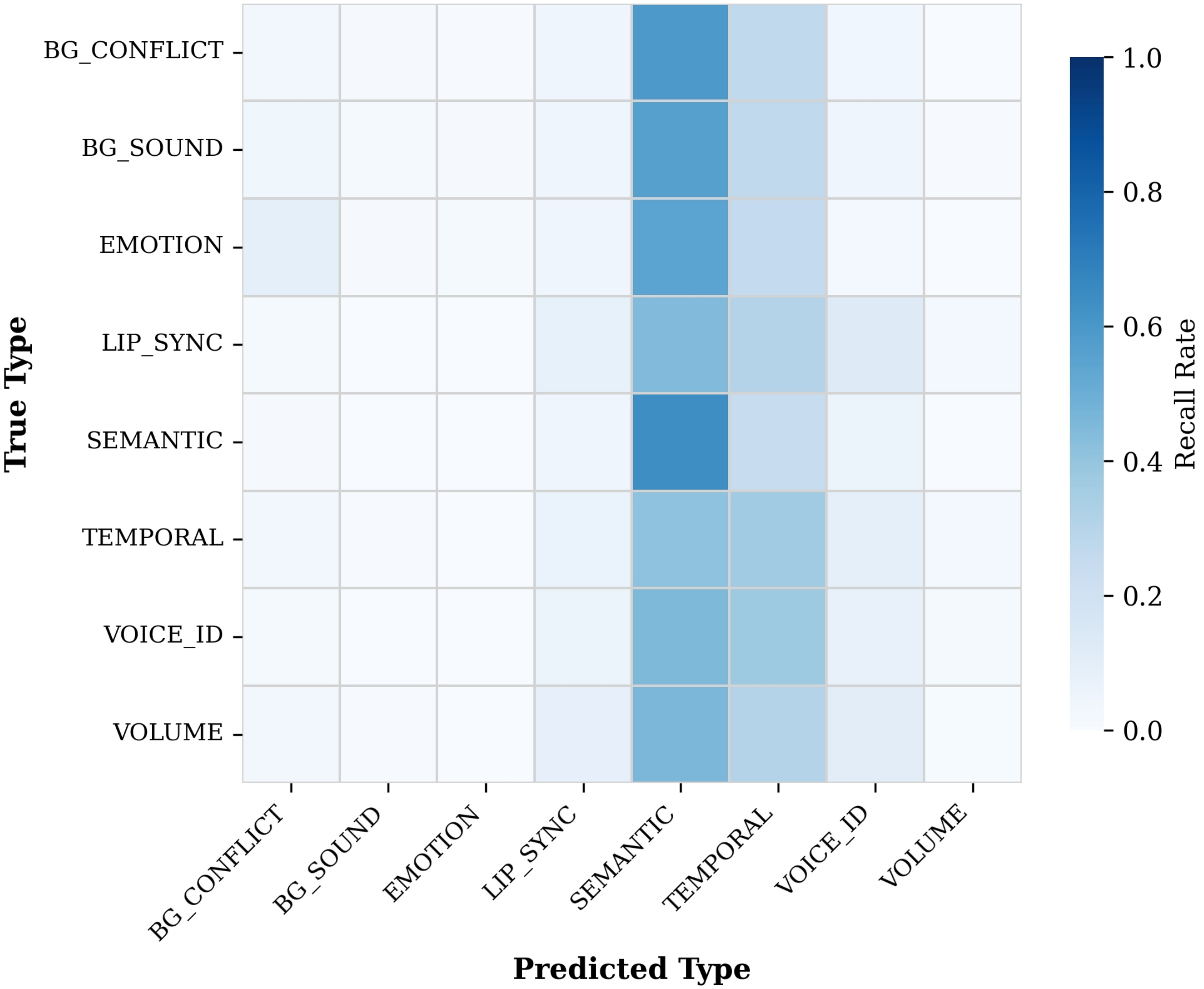}
        \\[-0.5em]
        {\small (g) Hunyuan}
    \end{minipage}\hfill
    \begin{minipage}[t]{0.24\linewidth}
        \centering
        \includegraphics[width=\linewidth]{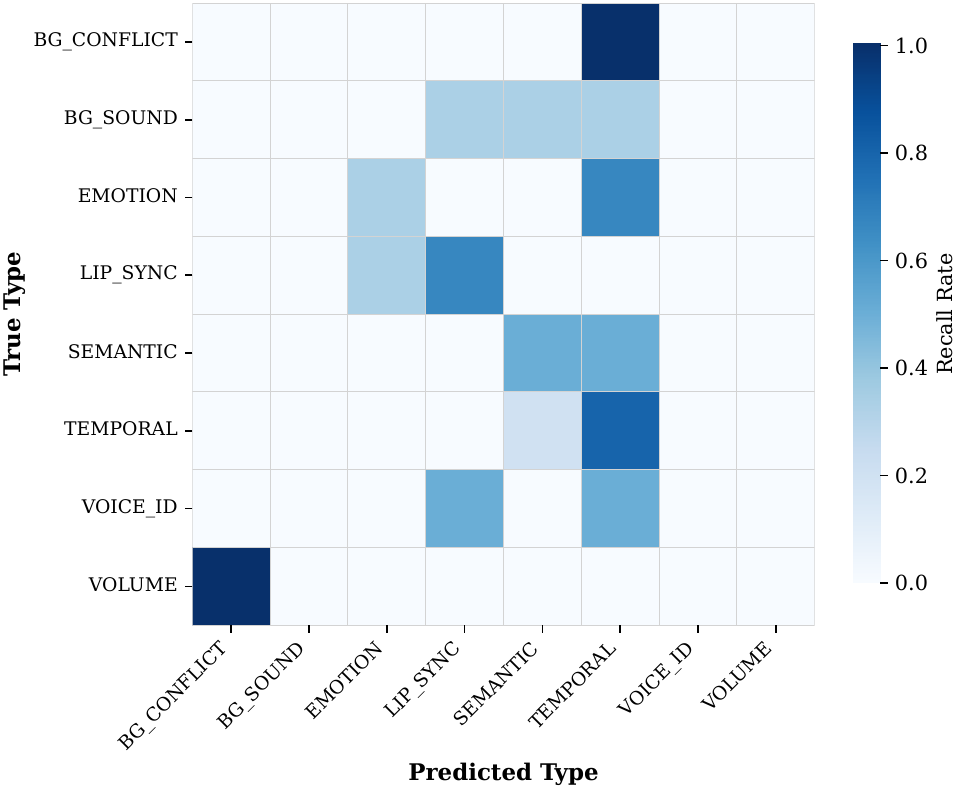}
        \\[-0.5em]
        {\small (h) SALMONN}
    \end{minipage}

    \caption{Confusion matrices on segment-level 8-category inconsistency classification: (top row) closed-source models and ours, (bottom row) open-source models.}
    \label{fig:segment_confusion_compact}
\end{figure}

\subsection{In-depth Analysis}\label{sec:exp_segment_confusion}

\noindent\textbf{Segment-level Task.}
To further understand model capabilities on segment-level tasks, we analyze from two perspectives: (1) recall vs. false positive rate, and (2) fine-grained classification ability.

\begin{figure}[t]
    \centering
    \begin{subfigure}[t]{0.5\linewidth}
        \centering
        \includegraphics[width=\linewidth]{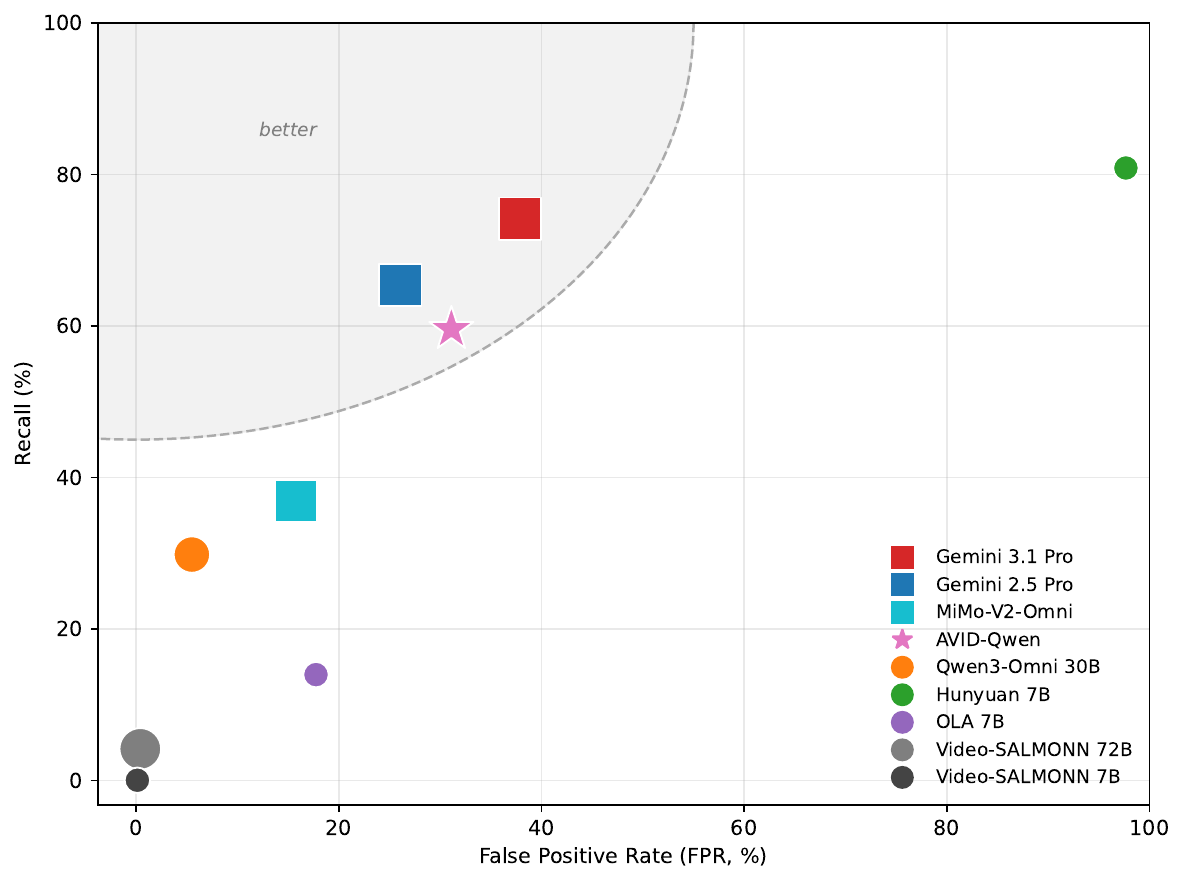}
        \caption{Recall vs. False Positive Rate}
        \label{fig:detection_scatter_a}
    \end{subfigure}\hfill
    \begin{subfigure}[t]{0.5\linewidth}
        \centering
        \includegraphics[width=\linewidth]{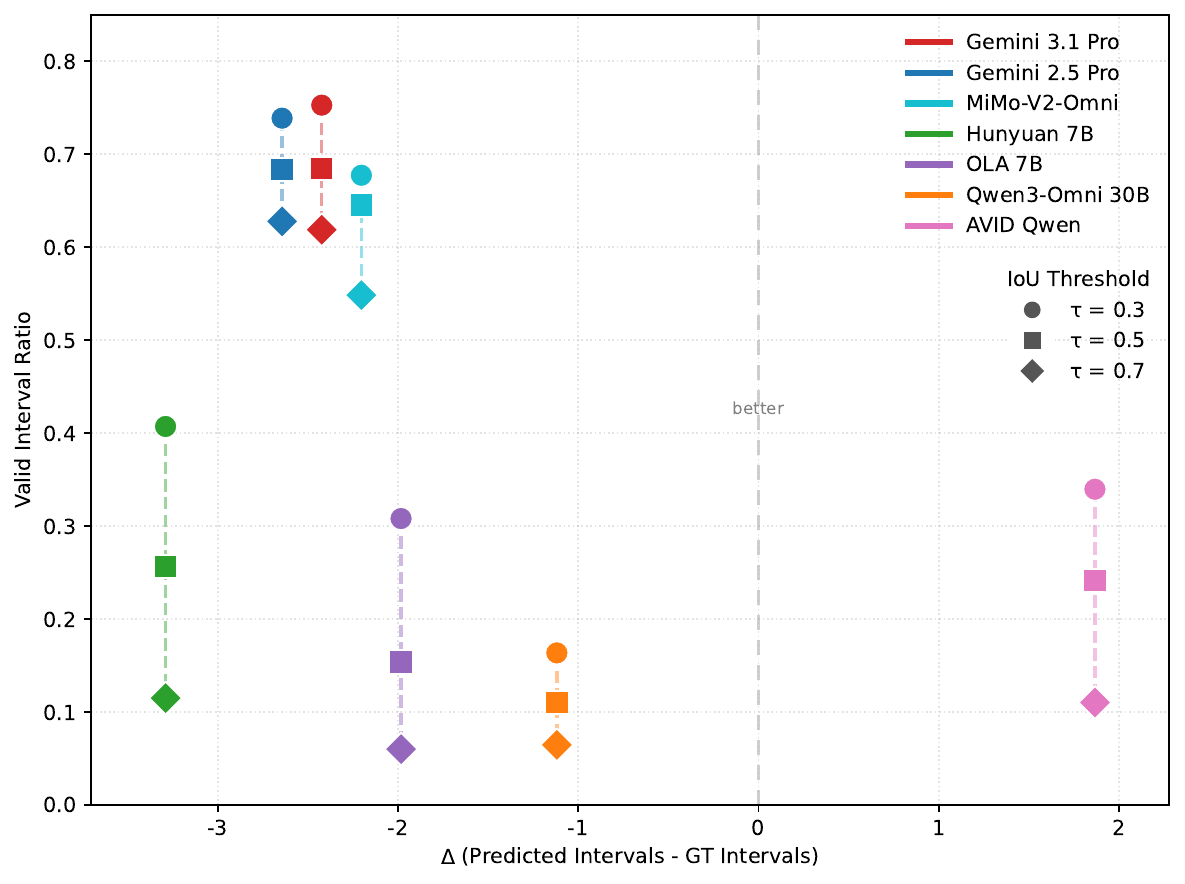}
        \caption{Grounding Quantity vs. Quality}
        \label{fig:detection_scatter_b}
    \end{subfigure}
    \caption{Model behavior analysis on full-video temporal grounding: (a) Recall vs. False Positive Rate — upper-left indicates balanced detection; (b) Grounding Quantity vs. Quality — x-axis closer to zero and y-axis higher indicates better performance.}
    \label{fig:detection_scatter}
\end{figure}

\textbf{Recall vs. False Positive Rate.} Figure~\ref{fig:detection_scatter_a} presents the recall-FPR scatter plot. The upper-left region represents the ideal zone where models achieve high recall while maintaining low false positive rate — indicating a balanced detection capability. Our analysis reveals that only Gemini models initially occupy this zone. After fine-tuning with AVID training data, AVID-Qwen successfully enters this region, demonstrating substantially improved detection balance. In contrast, other open-source models tend to cluster toward the extremes: either over-predicting inconsistencies (high recall, high FPR) or under-predicting (low recall, low FPR), revealing their difficulty in achieving balanced detection.

\textbf{Fine-grained Classification.} Figure~\ref{fig:segment_confusion_compact} presents confusion matrices for 8-category inconsistency classification. A pronounced diagonal structure (i.e., higher values along the diagonal than off-diagonal entries) indicates effective fine-grained detection across all categories. Gemini models and MiMo-V2-Omni exhibit strong diagonal dominance, demonstrating balanced classification ability across all 8 categories. After fine-tuning, AVID-Qwen also displays pronounced diagonal structure — this confirms substantially improved fine-grained detection capability. In contrast, other open-source models (OLA, ARC-Hunyuan-Video, Qwen3-Omni) lack diagonal dominance, indicating they can only distinguish 1-2 categories (e.g., OLA focuses on "lip sync" and "semantic", while Qwen3-Omni mainly detects "background sound" and "semantic"). This reveals that most open-source models lack sufficient capability for fine-grained inconsistency classification.

\noindent\textbf{Full-video Task.}
To better understand model behavior in temporal grounding, we analyze the trade-off between prediction quantity and prediction quality. Specifically, we measure: (1) the difference between the average number of predicted intervals and the average number of ground-truth intervals (x-axis), where values closer to zero indicate more accurate quantity estimation; and (2) the proportion of valid predictions (y-axis), defined as the ratio of predicted intervals whose IoU with ground-truth exceeds a threshold (0.3, 0.5, or 0.7), where higher values indicate better grounding precision.

As shown in Figure~\ref{fig:detection_scatter_b}, Gemini models and MiMo-V2-Omni cluster in the upper-left region, indicating a conservative but precise grounding strategy: they predict fewer intervals than ground-truth on average, but a large proportion of their predictions are correct. In contrast, open-source models exhibit weaker and less stable behavior. For example, Qwen3-Omni predicts a number of intervals close to the ground-truth count, but with very low precision, while OLA and ARC-Hunyuan-Video predict fewer intervals with only moderate validity.
Notably, AVID-Qwen predicts more intervals than the ground-truth on average, which lowers the proportion-based precision score because the denominator becomes larger. However, this does not indicate weaker grounding ability. Instead, it suggests that AVID-Qwen is able to identify substantially more valid inconsistency intervals, while also attempting more challenging cases that are difficult to localize. This behavior is consistent with its strong grounding performance in Table~\ref{tab:model_comparison}, where it achieves the best overall results. Taken together, these findings suggest that AVID-Qwen shifts the model from a conservative grounding strategy toward a more comprehensive one, improving interval coverage without sacrificing overall grounding effectiveness.

\begin{table}[htbp]
    \centering
    \footnotesize
    \setlength{\tabcolsep}{3pt}
    \caption{Ablation study on training stages.}
    \label{tab:ablation_training_stages}
    \begin{tabular}{l|c c cc|c cc c c}
        \toprule
        \multirow{3}{*}{\textbf{Model}} & \multicolumn{4}{c|}{\textbf{Segment}} & \multicolumn{5}{c}{\textbf{Full-Video}}\\
        \cmidrule{2-10}
        & Detect & Cla. & \multicolumn{2}{c|}{\textbf{Reasoning}} &  Detect & \multicolumn{2}{c}{\textbf{Grounding}} & reasoning& Dense\\
        \cmidrule{2-10}
        & Acc & Acc & B & R & Acc & R@0.5 & mIoU & R & $S_M$ \\
        \midrule
        Base & 55.7 & 55.0 & 2.2 & 21.7 & 75.3 & 8.3 & 9.1 & 3.90 & 1.48 \\
        Segment FT & 61.5 & 58.5 & 5.8 & 26.2 & 62.5 & 4.6 & 5.1 & 2.64 & 1.25 \\
        Full-Video FT & 57.5 & 51.2 & 4.0 & 20.4 & 76.4 & 31.6 & 31.7 & 14.6 & 6.11 \\
        AVID-Qwen & 61.3 & 55.5 & 6.2 & 26.5 & 78.2 & 39.2 & 36.1 & 15.2 & 7.47 \\
        \bottomrule
    \end{tabular}
\end{table}

\subsection{Ablation Study}\label{sec:exp_ablation}

We analyze the effect of training data by comparing four models: \textbf{base Qwen3-Omni}, \textbf{Segment FT} (trained on segment data only), \textbf{Full-Video FT} (trained on full-video data only), and \textbf{AVID-Qwen} (two-stage training on both). Quantitative results are summarized in Table~\ref{tab:ablation_training_stages}.
\textbf{Segment FT vs. Base.} Training on segment data mainly benefits segment-level tasks. For example, segment detection improves from 55.7\% to 61.5\%. However, it hurts full-video understanding, as reflected by the drop of full-video R@0.5 from 8.3\% to 4.6\%. This suggests that segment-only training improves fine-grained local prediction but weakens long-range temporal understanding.
\textbf{Full-Video FT vs. Base.} In contrast, training on full-video data mainly benefits full-video tasks. For example, full-video R@0.5 increases from 8.3\% to 31.6\%. However, segment-level classification drops from 55.0\% to 51.2\%, indicating that full-video supervision improves global temporal reasoning but is less effective for fine-grained segment discrimination.
\textbf{AVID-Qwen vs. Both.} Two-stage training achieves the best balance between the two settings. It preserves strong segment-level performance while further improving full-video grounding, reaching 39.2\% R@0.5. This confirms that segment-level and full-video supervision are complementary, and that combining them yields the strongest overall performance.

Notably, classification accuracy shows limited improvement (55.0\% → 55.5\% for AVID-Qwen vs. Base), suggesting that reasoning and classification capabilities may not be well-aligned — the model can generate accurate reasoning but struggles to map it to the correct inconsistency category. Improving classification likely requires joint training with explicit reasoning-type alignment objectives.  We leave these improvements for future work.

\section{Conclusion}

We present AVID, the first large-scale benchmark for audio-visual \emph{inconsistency} understanding in long-form videos. Our work makes three primary contributions: (1) a scalable construction pipeline comprising temporal segmentation, agent-driven strategy planning, and five specialized injectors for diverse inconsistency injection; (2) a comprehensive benchmark with 11.2K long videos, 39.4K annotated inconsistency events, and 78.7K segment-level clips, supporting evaluation across detection, temporal grounding, classification, and reasoning tasks; and (3) extensive evaluations of state-of-the-art omni-models and a strong fine-tuned baseline (AVID-Qwen) demonstrating substantial improvements.

Our experimental findings reveal critical limitations in current omni-models: (i) temporal grounding remains a significant challenge, with even the best closed-source model (Gemini 3.1 Pro) achieving only 26.2\% mIoU; (ii) fine-grained classification and reasoning capabilities are insufficient, particularly for open-source models; and (iii) reasoning and classification are not well-aligned, suggesting the need for joint training with explicit reasoning-type alignment objectives. AVID-Qwen achieves the best overall performance, surpassing all compared models in temporal grounding (mIoU: 36.1\%) and holistic understanding (SODA-m: 7.47), validating the effectiveness of both our benchmark and the two-stage fine-tuning strategy.

We anticipate AVID will serve as a rigorous testbed for advancing trustworthy, physics-aware omni-modal AI systems. Future work includes exploring reasoning-type alignment training objectives, expanding inconsistency categories, and developing models with stronger cross-modal reasoning capabilities.


\bibliographystyle{ACM-Reference-Format}
\bibliography{bib/references}

\appendix
\section{Appendix Overview}\label{sec:appendix_overview}

This appendix provides implementation-level details that are intentionally omitted from the main paper for space reasons. We organize the content into four main sections: (A) Injector implementation details for benchmark construction, (B) Model architecture and training hyper-parameters for \baselinename, (C) Evaluation protocol including metrics and model configurations, and (D) Prompt templates used in both construction and testing.

\section{Injector Implementations and Hyper-Parameters}\label{sec:appendix_injectors}\label{sec:injector_details}

This section provides detailed implementation specifications for the five injectors used in the \modelname construction pipeline.

\subsection{Global Pipeline Interface}\label{sec:appendix_injectors_interface}
Each injector follows a unified interface:
\begin{equation}
\mathcal{I}_k: (V, A, M) \rightarrow (V', A', \Delta, \Theta_k),
\end{equation}
where $V/A$ are original video/audio streams, $M$ denotes strategy metadata, $\Delta$ is generated annotation delta, and $\Theta_k$ is injector-specific configuration.

\subsection{Injector Overview Tables}\label{sec:appendix_injector_tables}

\begin{table*}[t]
     \centering
     \caption{Overview of the five injection techniques in \modelname.}
     \label{tab:injectors}
     \resizebox{\textwidth}{!}{
     \begin{tabular}{clll}
     \toprule
     \textbf{\#} & \textbf{Category} & \textbf{Core Mechanism} & \textbf{Key Tools} \\
     \midrule
     1 & TemporalInjector (Temporal Shift) & Audio time-axis offset ($\pm$0.5--3.0\,s) & FFmpeg (PCM extraction + temporal remapping) \\
     2 & SemanticInjector (Semantic Contradiction) & Replace speech with contradictory TTS while preserving background & Demucs + Qwen3-TTS + FFmpeg (amix) \\
     3 & IdentityInjector (Vocal Identity Conflict) & Modify pitch/formant/brightness for identity mismatch & FFmpeg (asetrate, atempo, equalizer, tremolo) \\
     4 & SpatialInjector (Spatial Acoustics) & Distance-based volume attenuation (near/far field) & FFmpeg (volume envelope) \\
     5 & BackgroundInjector (Environmental/Affective Conflict) & Replace background audio with scene-contradictory ambience or affective music & Demucs + Background Library + FFmpeg \\
     \bottomrule
     \end{tabular}
     }
\end{table*}

\begin{table*}[t]
\centering
\footnotesize
\caption{Class-conditional mapping between segment classes and applicable inconsistency categories.}
\label{tab:class_mapping}
\begin{tabular}{llll}
\toprule
\textbf{Segment Class} & \textbf{Inconsistency Category} & \textbf{Injector} & \textbf{Description} \\
\midrule
\multirow{4}{*}{Class~1 (Active Speaker)} & VOICE\_IDENTITY & IdentityInjector & Timbre/identity conflict (age/gender/voice-profile shift) \\
& VOLUME\_FLUCTUATION & SpatialInjector & Loudness changes without corresponding visual movement \\
& LIP\_SYNC & SemanticInjector & Mismatch between generated speech and lip motion \\
& TEMPORAL\_SHIFT & TemporalInjector & Audio leads/lags video by 0.5--2.0\,s \\
\midrule
\multirow{2}{*}{Class~2 (Voiceover)} & BACKGROUND\_CONFLICT & BackgroundInjector & Narration/music conflicts with visual scene context \\
& SEMANTIC\_DIVERGENCE & SemanticInjector & Spoken content contradicts visual semantics \\
\midrule
\multirow{2}{*}{Class~3 (Scenic)} & EMOTION\_MISMATCH & BackgroundInjector & Background mood conflicts with scene affect \\
& BACKGROUND\_SOUND & BackgroundInjector & Background sound conflicts with scene environment \\
\bottomrule
\end{tabular}
\end{table*}

\subsection{Injector-1: Temporal Shift Injector}\label{sec:appendix_injector1}
\textbf{Purpose.} Create temporal asynchrony between visible articulation and speech by applying audio time-axis offset.

\textbf{Implementation details.}
\begin{itemize}[leftmargin=*]
    \item Extract audio stream using FFmpeg with PCM format (48kHz, stereo) for precise sample-level manipulation.
    \item Apply temporal offset: positive values delay audio (audio lags behind video), negative values advance audio (audio leads video).
    \item For delay: prepend silence and truncate audio end. For advance: skip audio beginning and append silence.
    \item All timestamps are rounded to millisecond precision to avoid FFmpeg precision issues.
    \item Use \texttt{atrim} filter for sample-accurate audio cutting and \texttt{concat} demuxer for seamless segment merging.
\end{itemize}

\textbf{Key parameters.}
\begin{itemize}[leftmargin=*]
    \item Shift range: $\delta t \in [0.5, 3.0]$ seconds (positive = audio lags, negative = audio leads).
    \item Audio format: PCM 16-bit, 48kHz sample rate, stereo channels.
    \item Output encoding: AAC at 192kbps for final video.
\end{itemize}

\subsection{Injector-2: Semantic Divergence Injector}\label{sec:appendix_injector2}
\textbf{Purpose.} Replace original speech with semantically contradictory TTS-generated speech while preserving background audio.

\textbf{Implementation details.}
\begin{itemize}[leftmargin=*]
    \item \textbf{Step 1 - Audio Separation}: Use Demucs (htdemucs model) to separate the original audio into vocals and background (no\_vocals) tracks.
    \item \textbf{Step 2 - TTS Generation}: Generate contradictory speech using Qwen3-TTS with voice cloning. The original segment audio serves as the reference for voice cloning, ensuring speaker identity preservation while changing content.
    \item \textbf{Step 3 - Duration Matching}: Adjust TTS audio duration using FFmpeg \texttt{atempo} filter. If TTS is too short, slow down (atempo 0.7--0.9); if too long, speed up (atempo 1.1--1.3).
    \item \textbf{Step 4 - Mixing}: Combine the new TTS vocals with the preserved background audio using FFmpeg \texttt{amix} filter.
\end{itemize}

\textbf{Key parameters.}
\begin{itemize}[leftmargin=*]
    \item Voice cloning model: Qwen3-TTS-12Hz-1.7B-Base with x-vector extraction.
    \item Audio separation model: Demucs htdemucs (two-stems mode: vocals/no\_vocals).
    \item Volume mixing ratio: vocals = 1.0, background = 0.8.
    \item Speed adjustment range: atempo $\in [0.7, 1.3]$.
    \item Contradictory text length guidelines: 5--10s segment $\rightarrow$ 15--25 words; 10--15s $\rightarrow$ 25--35 words; 15--30s $\rightarrow$ 35--50 words.
\end{itemize}

\subsection{Injector-3: Identity Modification Injector}\label{sec:appendix_injector3}
\textbf{Purpose.} Modify vocal characteristics to create speaker identity mismatch (e.g., gender swap, age change) while preserving speech content.

\textbf{Implementation details.}
\begin{itemize}[leftmargin=*]
    \item \textbf{Pitch Shifting}: Use FFmpeg \texttt{asetrate} to change pitch by semitones, followed by \texttt{atempo} to restore original duration.
    \item \textbf{Formant Adjustment}: Simulate vocal tract length changes using equalizer filters. Higher formant shift (>1.0) for female voice, lower (<1.0) for male voice.
    \item \textbf{Brightness Control}: Apply high-frequency or low-frequency emphasis using \texttt{equalizer} filter.
    \item \textbf{Age Effects}: Add \texttt{tremolo} filter (f=4.5Hz, d=0.3) for elderly voice simulation.
    \item \textbf{Physique Effects}: Add \texttt{bass} boost and low-frequency \texttt{equalizer} for larger body simulation.
\end{itemize}

\textbf{Voice transformation presets.}
\begin{itemize}[leftmargin=*]
    \item \textbf{Female}: pitch +6 semitones, formant 1.15, brightness +0.3
    \item \textbf{Female\_Young}: pitch +8 semitones, formant 1.2, brightness +0.5
    \item \textbf{Female\_Old}: pitch +5 semitones, formant 1.1, brightness +0.1, tremolo enabled
    \item \textbf{Male}: pitch -6 semitones, formant 0.85, brightness -0.2
    \item \textbf{Male\_Deep}: pitch -9 semitones, formant 0.75, brightness -0.4, bass boost enabled
    \item \textbf{Child}: pitch +10 semitones, formant 1.25, brightness +0.6, speed 1.1x
    \item \textbf{Elder}: pitch -5 semitones, formant 0.9, brightness -0.2, tremolo enabled
\end{itemize}

\subsection{Injector-4: Spatial Mismatch Injector}\label{sec:appendix_injector4}
\textbf{Purpose.} Inject spatially inconsistent audio cues by simulating distance changes of the sound source (volume attenuation), while the visual context remains unchanged. This creates the perception that the sound source is moving closer or farther away without corresponding visual movement.

\textbf{Implementation details.}
\begin{itemize}[leftmargin=*]
    \item Extract the target audio segment from the video using FFmpeg.
    \item Apply volume envelope modulation to simulate distance changes: near-field (louder) to far-field (quieter).
    \item Use smooth transitions (fade-in/fade-out) to avoid abrupt volume changes.
    \item Merge the modified audio back into the original video while preserving the visual stream.
\end{itemize}

\textbf{Key parameters.}
\begin{itemize}[leftmargin=*]
    \item Volume range: 0.01 (far-field, nearly silent) to 1.0 (near-field, original volume).
    \item Fade duration: 50\% of segment duration for gradual transition, remaining 50\% holds the final volume.
    \item Direction modes: ``away'' (1.0 $\rightarrow$ 0.01, simulating moving away) or ``toward'' (0.01 $\rightarrow$ 1.0, simulating approaching).
    \item Audio format: AAC at 192kbps, 48kHz sample rate, stereo channels.
\end{itemize}

\subsection{Injector-5: Background Injector}\label{sec:appendix_injector5}
\textbf{Purpose.} Replace the environmental background audio with contradictory ambience or affective music, while preserving the foreground speech if available. This generates inconsistencies related to the environment or emotional context.

\textbf{Implementation details.}
\begin{itemize}[leftmargin=*]
    \item Extract the target audio segment from the video using FFmpeg.
    \item Use Demucs (htdemucs model) to separate vocals (foreground speech) from background audio.
    \item Select replacement audio from our sound effects library based on strategy parameters:
    \begin{itemize}
        \item \textbf{Background sound type}: city\_traffic, rain, thunder, wind, birds, ocean\_waves, dogs, cats, car\_horn, siren, subway, train, construction, factory, heavy\_machinery.
        \item \textbf{Emotion type}: happy, sad, peaceful, exciting, tense music tracks.
    \end{itemize}
    \item Adjust the selected audio duration to match the target segment (via truncation or looping).
    \item Mix the separated vocals with the new background audio at appropriate volume ratios (vocals: 1.0, background music: 0.5, background sound: 0.6).
    \item Merge the mixed audio back into the original video.
\end{itemize}

\textbf{Sound effects library.}
Our sound effects library contains 20 categories with multiple audio files each:
\begin{itemize}
    \item \textbf{Urban}: city\_traffic, construction, siren, subway, train, car\_horn
    \item \textbf{Nature}: rain, thunder, wind, birds, ocean\_waves
    \item \textbf{Animals}: dogs, cats
    \item \textbf{Machinery}: heavy\_machinery, factory
    \item \textbf{Music}: music\_happy, music\_sad, music\_peaceful, music\_excited, music\_tense
\end{itemize}

\textbf{Key parameters.}
\begin{itemize}[leftmargin=*]
    \item Background sound type or emotion category (selected by strategy agent).
    \item Volume ratio: 0.5--0.6 for background, 1.0 for foreground vocals.
    \item Duration matching: truncation for long audio, looping or concatenation for short audio.
\end{itemize}

\section{Model Architecture and Training Details}\label{sec:appendix_model_train}\label{sec:appendix_model_arch}

This section provides detailed specifications of the \baselinename model architecture and training hyper-parameters.

\subsection{Model Overview}
Qwen3-Omni is an omni-modal large language model with approximately 30 billion total parameters (with 3 billion activated parameters). The model consists of three core components:

\begin{itemize}
    \item \textbf{Vision Encoder}: Processes input video frames into visual tokens. The encoder is frozen during fine-tuning to preserve pre-trained visual understanding capabilities.
    \item \textbf{Audio Encoder}: Extracts acoustic features directly from the audio stream embedded in the video. The audio encoder is also frozen during fine-tuning.
    \item \textbf{LLM Backbone}: A decoder-only transformer that performs multimodal reasoning over concatenated visual tokens, audio tokens, and text tokens.
    \item \textbf{Aligner Module}: Bridges the vision encoder outputs to the LLM embedding space. This module is frozen during fine-tuning.
\end{itemize}

\subsection{Input Preprocessing}
The input preprocessing pipeline follows the default Qwen3-Omni configuration:

\begin{itemize}
    \item \textbf{Video Frame Sampling}: Videos are sampled at 12 frames per second (FPS).
    \item \textbf{Resolution}: Each frame is resized to a maximum of 50,176 pixels while preserving aspect ratio. The maximum sequence length for visual tokens is set to 1,003,520 pixels.
    \item \textbf{Audio Extraction}: Audio is directly extracted from the video file using the decord reader. The audio sampling rate follows the model's default configuration.
    \item \textbf{Max Sequence Length}: The maximum token sequence length is set to 8,192.
\end{itemize}

The model accepts a conversation format where the user message contains interleaved video and text. The video is specified using the \texttt{<video>} token, and the multimodal processor handles frame sampling and audio extraction automatically.

\subsection{Training Hyper-Parameters}\label{sec:appendix_train_hparams}

We provide the complete training hyper-parameters for both fine-tuning stages. Table~\ref{tab:train_hparams} summarizes the configuration.

\begin{table}[h]
    \centering
    \caption{Training hyper-parameters for \baselinename two-stage fine-tuning.}
    \label{tab:train_hparams}
    \resizebox{\linewidth}{!}{
    \begin{tabular}{lcc}
        \toprule
        \textbf{Parameter} & \textbf{Stage 1 (Segment)} & \textbf{Stage 2 (Full-Video)} \\
        \midrule
        Base Model & \multicolumn{2}{c}{Qwen3-Omni-30B-A3B-Instruct} \\
        Training Type & \multicolumn{2}{c}{LoRA} \\
        LoRA Rank & \multicolumn{2}{c}{8} \\
        LoRA Alpha & \multicolumn{2}{c}{32} \\
        LoRA Target Modules & \multicolumn{2}{c}{all-linear} \\
        Frozen Modules & \multicolumn{2}{c}{Vision Encoder, Aligner} \\
        \midrule
        Learning Rate & \multicolumn{2}{c}{1e-4} \\
        Optimizer & \multicolumn{2}{c}{AdamW (bfloat16)} \\
        Warmup Ratio & \multicolumn{2}{c}{0.05} \\
        \midrule
        Per Device Batch Size & \multicolumn{2}{c}{2} \\
        Gradient Accumulation Steps & \multicolumn{2}{c}{4} \\
        Effective Batch Size & \multicolumn{2}{c}{48 (6 GPUs × 2 × 4)} \\
        \midrule
        Training Epochs & 2 & 10 \\
        Max Sequence Length & \multicolumn{2}{c}{8192} \\
        \midrule
        FPS Max Frames & \multicolumn{2}{c}{12} \\
        Video Max Pixels & \multicolumn{2}{c}{50,176} \\
        \midrule
        Hardware & \multicolumn{2}{c}{6 × NVIDIA A100 80GB} \\
        Distributed Training & \multicolumn{2}{c}{DeepSpeed ZeRO-3} \\
        \bottomrule
    \end{tabular}
    }
\end{table}

The training uses DeepSpeed ZeRO-3 optimization for memory-efficient distributed training across 6 A100 GPUs. Gradient checkpointing is enabled to reduce memory usage during backpropagation. The model is trained with the SWIFT fine-tuning framework.

\section{Evaluation Protocol}\label{sec:appendix_eval}\label{sec:appendix_metrics}\label{sec:appendix_llm_config}

This section provides detailed metric definitions, computation protocols, and model testing configurations used in our experiments.

\subsection{Notation}\label{sec:appendix_metrics_notation}
For a sample $i$, let $y_i \in \{0,1\}$ denote the binary inconsistency label, and $\hat{y}_i$ the model prediction. For temporal grounding, let the predicted interval be $\hat{g}_i=[\hat{t}_{s},\hat{t}_{e}]$ and the ground-truth interval be $g_i=[t_s,t_e]$.

\subsection{Binary Detection Metrics}\label{sec:appendix_metrics_detection}
\textbf{Accuracy.}
\begin{equation}
\mathrm{Acc}=\frac{1}{N}\sum_{i=1}^{N}\mathbb{I}(\hat{y}_i=y_i).
\end{equation}

\textbf{Precision, Recall, F1 (optional reporting).}
\begin{equation}
\mathrm{Precision}=\frac{TP}{TP+FP},
\end{equation}

\begin{equation}
    \mathrm{Recall}=\frac{TP}{TP+FN},
\end{equation}

\begin{equation}
    \mathrm{F1}=\frac{2\cdot \mathrm{Precision}\cdot \mathrm{Recall}}{\mathrm{Precision}+\mathrm{Recall}}.
\end{equation}

\textbf{False Positive Rate (for scatter analysis).}
\begin{equation}
\mathrm{FPR}=\frac{FP}{FP+TN}.
\end{equation}

\subsection{Category Classification Metrics}\label{sec:appendix_metrics_type}
\textbf{Top-1 category accuracy.}
\begin{equation}
\mathrm{Acc}_{\mathrm{type}}=\frac{1}{N_{\mathrm{inc}}}\sum_{i=1}^{N_{\mathrm{inc}}}\mathbb{I}(\hat{c}_i=c_i),
\end{equation}
where $c_i$ and $\hat{c}_i$ are ground-truth and predicted inconsistency categories.

We report Top-1 accuracy as the primary classification metric. Macro-F1 and Weighted-F1 are not included in the main results.

\subsection{Text Reasoning Metrics}\label{sec:appendix_metrics_reasoning}
We compute BLEU-4, ROUGE-L, and METEOR between model-generated explanations $\hat{r}$ and reference explanations $r$ under standardized tokenization and lower-casing. If multiple references exist, we follow the official maximum-over-references protocol. All scores are multiplied by 100 for percentage-scale reporting.

\textbf{BLEU-4.}
BLEU (Bilingual Evaluation Understudy) measures n-gram precision between hypothesis and reference. We use BLEU-4 with smoothing to handle short sentences:
\begin{equation}
\mathrm{BLEU\text{-}4} = \mathrm{BP} \cdot \exp\left(\sum_{n=1}^{4} w_n \log p_n\right),
\end{equation}
where $p_n$ is the modified n-gram precision, $w_n = \frac{1}{4}$ is uniform weight, and BP is the brevity penalty:
\begin{equation}
\mathrm{BP} = \begin{cases} 1 & \text{if } |\hat{r}| > |r| \\ \exp(1 - |r|/|\hat{r}|) & \text{otherwise} \end{cases}.
\end{equation}

\textbf{ROUGE-L.}
ROUGE-L measures the longest common subsequence (LCS) between hypothesis and reference:
\begin{equation}
R_{\mathrm{lcs}} = \frac{\mathrm{LCS}(\hat{r}, r)}{|r|}, \quad P_{\mathrm{lcs}} = \frac{\mathrm{LCS}(\hat{r}, r)}{|\hat{r}|},
\end{equation}
\begin{equation}
\mathrm{ROUGE\text{-}L} = F_{\mathrm{lcs}} = \frac{(1+\beta^2) R_{\mathrm{lcs}} P_{\mathrm{lcs}}}{R_{\mathrm{lcs}} + \beta^2 P_{\mathrm{lcs}}},
\end{equation}
where $\beta$ is set to favor recall (typically $\beta=1.2$). We report the F-measure $F_{\mathrm{lcs}}$.

\textbf{METEOR.}
METEOR (Metric for Evaluation of Translation with Explicit ORdering) combines unigram precision and recall with a penalty for fragmentation:
\begin{equation}
P_m = \frac{|m|}{|\hat{r}|}, \quad R_m = \frac{|m|}{|r|},
\end{equation}
\begin{equation}
F_{\mathrm{mean}} = \frac{10 P_m R_m}{R_m + 9 P_m},
\end{equation}
\begin{equation}
\mathrm{METEOR} = F_{\mathrm{mean}} \cdot (1 - \gamma \cdot \mathrm{frag}^\theta),
\end{equation}
where $|m|$ is the number of matched unigrams (including stems, synonyms, and paraphrases), $\mathrm{frag} = \frac{\text{\#chunks}}{|m|}$ is the fragmentation penalty, and $\gamma=0.5$, $\theta=3$ are default parameters.

\subsection{Temporal Grounding Metrics}\label{sec:appendix_metrics_grounding}
\textbf{IoU for temporal intervals.}
\begin{equation}
\mathrm{IoU}(\hat{g}_i,g_i)=\frac{|\hat{g}_i \cap g_i|}{|\hat{g}_i \cup g_i|}.
\end{equation}

\textbf{Recall@1 at threshold $\alpha$.}
\begin{equation}
\mathrm{R@1}(\alpha)=\frac{1}{N_g}\sum_{i=1}^{N_g}\mathbb{I}(\mathrm{IoU}(\hat{g}_i,g_i)\ge\alpha),\quad \alpha\in\{0.3,0.5,0.7\}.
\end{equation}

\textbf{mIoU.}
\begin{equation}
\mathrm{mIoU}=\frac{1}{N_g}\sum_{i=1}^{N_g}\mathrm{IoU}(\hat{g}_i,g_i).
\end{equation}

\begin{table*}[t]
    \centering
    \footnotesize
    \caption{Detailed test-time configuration for all evaluated large omni-models.}
    \label{tab:appendix_model_config}
    \begin{tabular*}{\textwidth}{@{\extracolsep{\fill}} l l l c c c c c c @{} }
        \toprule
        \textbf{Model} & \textbf{Provider/API} & \textbf{Version/Checkpoint} & \textbf{Context Len} & \textbf{Temp.} & \textbf{Top-p} & \textbf{Max Tokens} & \textbf{Frames/Rate} & \textbf{Audio Setting} \\
        \midrule
        Gemini 2.5 Pro & Google Vertex AI & gemini-2.5-pro-preview-0506 & 1M & 0.3 & - & 4096 & auto & Yes \\
        Gemini 3.1 Pro & Google Vertex AI & gemini-3.1-pro-preview & 1M & 0.3 & - & 4096 & auto & Yes \\
        ARC-Hunyuan-Video & Local (HuggingFace) & TencentARC/ARC-Hunyuan-Video-7B & 32K & 0.7 & 0.9 & 2048 & 8fps & Yes \\
        Qwen3-Omni & Local (HuggingFace) & Qwen3-Omni-30B-A3B-Instruct & 8K & 0.3 & - & 2048 & 12fps & Yes \\
        OLA & Local & Ola-main/inference & 8K & 0.3 & - & 2048 & 12fps & Yes \\
        Video-SALMONN 2-7B & Local & Video-SALMONN-2-7B & 8K & 0.3 & - & 2048 & 16kHz & Yes \\
        Video-SALMONN 2-72B & Local & Video-SALMONN-2-72B & 8K & 0.3 & - & 2048 & 16kHz & Yes \\
        MiMo-V2-Omni & OpenRouter & xiaomi/mimo-v2-omni & 32K & 0.3 & - & 4096 & auto & Yes \\
        Ours-v1 & Local (Qwen3-Omni) & Qwen3-Omni-30B-A3B-Instruct + SFT & 8K & 0.3 & - & 2048 & 12fps & Yes \\
        Ours-v2 & Local (Qwen3-Omni) & Qwen3-Omni-30B-A3B-Instruct + SFT & 8K & 0.3 & - & 2048 & 12fps & Yes \\
        \bottomrule
    \end{tabular*}
\end{table*}

\subsection{Dense Full-Video Metric}\label{sec:appendix_metrics_dense}
\textbf{SODA-m.}
We adopt SODA-m (Story-based Dense Annotation with Multi-reference) as a holistic dense understanding metric for full-video prediction quality. SODA-m evaluates the quality of dense temporal predictions by considering both localization accuracy and caption quality jointly.

Given a set of predicted events $\hat{\mathcal{E}} = \{(\hat{g}_j, \hat{c}_j)\}_{j=1}^{M}$ and ground-truth events $\mathcal{E} = \{(g_i, c_i)\}_{i=1}^{N}$, where each event consists of a temporal interval $g$ and a caption $c$, SODA-m first computes an optimal bipartite matching using the Hungarian algorithm:
\begin{equation}
\pi^* = \arg\max_{\pi} \sum_{i=1}^{\min(N,M)} \mathrm{IoU}(g_i, \hat{g}_{\pi(i)}),
\end{equation}
where $\pi$ is a permutation mapping ground-truth events to predicted events.

For each matched pair $(i, \pi^*(i))$, the localization quality is measured by IoU, and the caption quality is measured by a text similarity function $S(\cdot, \cdot)$ (e.g., METEOR). The per-pair score is:
\begin{equation}
s_i = \mathrm{IoU}(g_i, \hat{g}_{\pi^*(i)}) \cdot S(c_i, \hat{c}_{\pi^*(i)}).
\end{equation}

The final SODA-m score aggregates over all ground-truth events, penalizing both missed detections (unmatched ground-truth) and false alarms (unmatched predictions):
\begin{equation}
\mathrm{SODA\text{-}m} = \frac{\sum_{i=1}^{|\mathcal{M}|} s_i}{N + (M - |\mathcal{M}|)},
\end{equation}
where $\mathcal{M}$ is the set of matched pairs with $\mathrm{IoU} > 0$, $N$ is the number of ground-truth events, and $(M - |\mathcal{M}|)$ counts false positive predictions.

\subsection{Unified Evaluation Protocol}\label{sec:appendix_metrics_protocol}
All metrics are reported in percentage scale ($\times 100$). For each model, we use a fixed decoding configuration with temperature set to 0.3 in the primary result table. Each sample is evaluated once (single run) as the low temperature setting ensures deterministic outputs with minimal variance.

\subsection{Compared Model Endpoints and Versions}\label{sec:appendix_llm_models}
Table~\ref{tab:appendix_model_config} summarizes detailed model versions and test-time settings.

\subsection{Reproducible Inference Settings}\label{sec:appendix_llm_inference}
\begin{itemize}[leftmargin=*]
    \item \textbf{Decoding strategy}: Temperature-based sampling with $T=0.3$ for all models. This low temperature setting produces near-deterministic outputs while allowing slight variation.
    \item \textbf{Video sampling policy}: Model-dependent as shown in Table~\ref{tab:appendix_model_config}. Gemini models use automatic frame selection; local models use fixed FPS (8--16 fps depending on model).
    \item \textbf{Audio preprocessing}: Audio is extracted directly from video files using FFmpeg. For API-based models (Gemini, MiMo), audio is embedded in the video file sent via base64 encoding. For local models, audio is processed at the model's native sample rate.
    \item \textbf{Prompt wrapping and role format}: All models use a single-turn user message containing the evaluation prompt and video. The prompt templates are provided in Figures~\ref{fig:multiturn_av_inconsistency_prompt} and~\ref{fig:fullvideo_multiturn_prompt}.
    \item \textbf{Failure handling}: API timeout is set to 600 seconds. Failed requests are logged but not retried to ensure reproducibility. Malformed outputs (e.g., missing required fields) are parsed using regex patterns to extract partial information when possible.
\end{itemize}

\subsection{Video-SALMONN 2 Detection Behavior Analysis}\label{sec:appendix_salmonn_analysis}
During our evaluation, we observed an unusual phenomenon: Video-SALMONN 2 (both 7B and 72B variants) consistently predicts "No inconsistency" for nearly all samples, resulting in detection accuracy close to random chance (~25\% on the balanced full-video test set).

To verify this is not a code bug, we conducted the following investigations:

\begin{enumerate}
    \item \textbf{Content understanding verification}: We prompted SALM\allowbreak ONN to describe what it sees and hears in the video without asking about inconsistency. The model provided accurate descriptions of both visual and audio content, confirming that the model can properly process the multimodal inputs.
    \item \textbf{Decoding parameter experiments}: We tested different decoding configurations including temperature = 0.3, 0.7, and 1.0, and verified that \texttt{do\_sample} was enabled. The prediction behavior remained unchanged across all settings.
    \item \textbf{Output format verification}: We examined the raw model outputs and confirmed they correctly parsed the "No" response for inconsistency detection.
\end{enumerate}

Based on these checks, we conclude that SALMONN's behavior is a genuine model characteristic rather than an implementation artifact. This observation suggests that SALMONN may have a strong prior toward positive-consistency predictions, possibly due to its training data distribution or model architecture design.

\section{Prompt Templates}\label{sec:appendix_prompts}

\subsection{Strategy-Agent Prompt Templates}\label{sec:appendix_prompts_strategy}
The complete prompt template for strategy planning is provided in Figure~\ref{fig:prompt_strategy_planning}. This prompt is used by the Strategy Planning Agent (Gemini 3.1 Pro) to analyze video segments and generate injection plans. The prompt includes:
\begin{itemize}[leftmargin=*]
    \item Input specification: video duration, number of valid segments, and temporal segmentation table with class labels.
    \item Segment class definitions: Class 1 (Active Speaker), Class 2 (Voiceover), and Class 3 (Scenic), with their applicable injection types.
    \item Output requirements: balanced injection selection, feasibility screening, and detailed inconsistency analysis.
    \item JSON output schema: structured format including start/end timestamps, class label, injection type, parameters, and reasoning.
\end{itemize}
\begin{figure*}[htbp]
\centering

\begin{tcolorbox}[
    width=\textwidth,
    colback=brown!5!white,
    colframe=black,
    boxrule=0.5pt,
    arc=0pt,
    left=7pt,
    right=7pt,
    top=9pt,
    bottom=9pt
]
\small
\setstretch{1.08}

\tcbox[
    colback=black,
    colframe=black,
    coltext=white,
    boxrule=0pt,
    arc=2pt,
    left=6pt,
    right=6pt,
    top=2pt,
    bottom=2pt
]{\texttt{\# Prompt for multi-turn audio-visual inconsistency detection in full videos}}

\vspace{0.6em}

Please carefully watch this full video (not a short segment), analyze its audio and visual content, and determine whether there is audio-visual inconsistency.

\vspace{0.5em}
\noindent\textbf{8 Inconsistency Categories (must choose from these)}

\vspace{0.25em}
\noindent\textbf{Class 1 (Active Speaker -- interview/dialogue, person speaking in video)}
\begin{itemize}[leftmargin=1.4em,itemsep=0.25em,topsep=0.35em]
    \item \texttt{TEMPORAL\_SHIFT}: Temporal offset, audio leads or lags behind video by 0.5--2s (e.g., see mouth moving but sound is delayed by 0.5s).
    \item \texttt{LIP\_SYNC}: Lip-sync mismatch, TTS-generated voice does not match lip movement (e.g., video shows a man speaking but voice sounds like a woman).
    \item \texttt{VOICE\_IDENTITY}: Voice identity conflict, speaker's voice changes abruptly (e.g., video shows an elderly person but voice is a child's).
    \item \texttt{VOLUME\_FLUCTUATION}: Volume conflict, person is still but volume fluctuates (e.g., person stands still but voice sounds like it is moving closer and further).
\end{itemize}

\noindent\textbf{Class 2 (Voiceover -- narration, no speaker in video)}
\begin{itemize}[leftmargin=1.4em,itemsep=0.25em,topsep=0.35em]
    \item \texttt{SEMANTIC\_DIVERGENCE}: Semantic inconsistency, TTS text contradicts video content (e.g., video shows food but narration talks about phones).
    \item \texttt{BACKGROUND\_CONFLICT}: Background sound conflict, narration/music contradicts video scene (e.g., video shows office but background has bar music).
\end{itemize}

\noindent\textbf{Class 3 (Scenic -- scenery/scene, no human voice)}
\begin{itemize}[leftmargin=1.4em,itemsep=0.25em,topsep=0.35em]
    \item \texttt{EMOTION\_MISMATCH}: Background music emotion mismatch, video is sad scene but has happy music (e.g., video shows funeral but music is joyful).
    \item \texttt{BACKGROUND\_SOUND}: Background sound conflict, video is scene A but audio is scene B (e.g., video shows forest but has city traffic sounds; video shows chopping but no chopping sound).
\end{itemize}

\vspace{0.5em}
\noindent\textbf{System instruction}

Please answer in the following exact format (each item on a new line):
\begin{enumerate}[leftmargin=1.5em,itemsep=0.25em,topsep=0.35em]
    \item \textbf{Is there any inconsistency in this video?} Answer: \texttt{Yes} or \texttt{No}.
    \item \textbf{If Yes, for each inconsistency, provide:} \texttt{from X.Xs to Y.Ys, reasoning} -- each on a separate line. If No, write \texttt{N/A}.
\end{enumerate}

\noindent\textbf{Example output format}
\begin{verbatim}
Yes
from 0.0s to 7.8s, The background music of sad feelings creates emotional conflict with the lively visual tone
from 15.5s to 20.3s, The sound of rain is injected but no rain is shown in the indoor scene
\end{verbatim}

\end{tcolorbox}

\caption{Prompt for multi-turn audio-visual inconsistency detection in full videos.}
\label{fig:fullvideo_multiturn_prompt}
\end{figure*}

\begin{figure*}[t]
\centering

\begin{tcolorbox}[
    width=\textwidth,
    colback=brown!5!white,
    colframe=black,
    boxrule=0.5pt,
    arc=0pt,
    left=7pt,
    right=7pt,
    top=9pt,
    bottom=9pt
]
\small
\setstretch{1.08}

\tcbox[
    colback=black,
    colframe=black,
    coltext=white,
    boxrule=0pt,
    arc=2pt,
    left=6pt,
    right=6pt,
    top=2pt,
    bottom=2pt
]{\texttt{\# Strategy planning prompt}}

\vspace{0.6em}

You are a professional expert in constructing audio-visual inconsistency data. Please select appropriate injection strategies for a video based on its temporal segmentation information.

The input video contains the following information: total duration \texttt{\{duration:.1f\}} seconds, \texttt{\{len(valid\_segments)\}} valid segments, and a temporal segmentation table. Please use the provided time ranges directly and do not estimate them yourself.

\vspace{0.35em}
\noindent
\textbf{Time range (duration) \textbar\ category}\\
\texttt{\{segments\_text\}}

\vspace{0.5em}

The segments are divided into three categories: \texttt{class\_1\_active\_speaker} (speech with visible speaker), \texttt{class\_2\_voiceover} (speech without visible speaker), and \texttt{class\_3\_scenic} (no speech, only background sound). The corresponding optional injection types are:
\begin{itemize}[leftmargin=1.3em,itemsep=0.25em,topsep=0.35em]
    \item \texttt{class\_1\_active\_speaker}: \texttt{VOICE\_IDENTITY}, \texttt{VOLUME\_FLUCTUATION}, \texttt{LIP\_SYNC}, \texttt{TEMPORAL\_SHIFT}
    \item \texttt{class\_2\_voiceover}: \texttt{BACKGROUND\_CONFLICT}, \texttt{SEMANTIC\_DIVERGENCE}
    \item \texttt{class\_3\_scenic}: \texttt{EMOTION\_MISMATCH}, \texttt{BACKGROUND\_SOUND}
\end{itemize}

\vspace{0.2em}
Please analyze the video and generate an injection plan under the following requirements:

\begin{enumerate}[leftmargin=1.5em,itemsep=0.25em,topsep=0.35em]
    \item \textbf{Segment selection.} Select appropriate segments for injection. Each selected segment must be at least 5 seconds and at most 30 seconds long. If a segment exceeds 30 seconds, subdivide it into smaller non-overlapping sub-segments, each treated as an independent injection plan.
    \item \textbf{Balance and preference.} Construct \texttt{\{min\_injections\}}--\texttt{\{max\_injections\}} injections depending on video length. Maintain class balance whenever possible by including at least one segment from each available class. Diversify injection types across selected segments. Prefer \texttt{TEMPORAL\_SHIFT} and \texttt{LIP\_SYNC}, and use \texttt{EMOTION\_MISMATCH} and \texttt{BACKGROUND\_SOUND} less frequently.
    \item \textbf{Inconsistency analysis.} For each selected segment, assign one valid injection type and provide a detailed explanation including visual description, audio description, inconsistency point, a fine-grained \texttt{dataset\_sub\_category}, and reasoning.
    \item \textbf{Output format.} Return the result in strict JSON format:
\end{enumerate}

\begin{verbatim}
{
  "injection_plans": [{
    "start": "...",
    "end": "...",
    "class_label": "...",
    "injection_type": "...",
    "injection_params": {...},
    "inconsistency_understanding": {
      "visual_description": "...",
      "audio_description": "...",
      "inconsistency_point": "..."
    },
    "dataset_sub_category": "...",
    "reasoning": "..."
  }],
  "summary": "..."
}
\end{verbatim}

Important constraints: selected segments must not overlap; \texttt{injection\_type} must be chosen from the predefined list; and \texttt{injection\_params} must follow the corresponding format. In particular, \texttt{TEMPORAL\_SHIFT} uses \texttt{\{"shift\_seconds": 1.0\}} (0.5--3.0 seconds; positive = audio leads, negative = audio lags); \texttt{LIP\_SYNC} uses \texttt{\{"text": "...", "voice\_type": "young\_female"\}}; \texttt{SEMANTIC\_DIVERGENCE} uses \texttt{\{"contradictory\_text": "...", "voice\_type": "young\_female"\}}, where text length should match clip duration (5--10 s: 15--25 words; 10--15 s: 25--35; 15--30 s: 35--50; 20+ s: 50--70); \texttt{VOICE\_IDENTITY} uses \texttt{\{"target\_voice": "Male"\}}; \texttt{VOLUME\_FLUCTUATION} uses \texttt{\{"direction": "away"\}} or \texttt{\{"direction": "toward"\}}; \texttt{BACKGROUND\_CONFLICT} and \texttt{BACKGROUND\_SOUND} use \texttt{\{"bg\_sound\_type": "city\_traffic"\}}; and \texttt{EMOTION\_MISMATCH} uses \texttt{\{"emotion": "happy"\}}.

\end{tcolorbox}

\caption{Prompt for strategy planning agent.}
\label{fig:prompt_strategy_planning}
\end{figure*}

\subsection{Evaluation Prompt Templates}\label{sec:appendix_prompts_eval}
We provide two evaluation prompt templates:

\begin{itemize}[leftmargin=*]
    \item \textbf{Segment-level evaluation} (Figure~\ref{fig:multiturn_av_inconsistency_prompt}): A multi-turn prompt for detecting inconsistency, classifying category, and providing reasoning explanation for short video segments.
    \item \textbf{Full-video evaluation} (Figure~\ref{fig:fullvideo_multiturn_prompt}): A prompt for detecting multiple inconsistencies in long videos, providing temporal grounding (start/end timestamps) and reasoning for each detected event.
\end{itemize}

Both prompts include the complete 8-category inconsistency taxonomy with examples and enforce structured output format for automated parsing.

\begin{figure*}[b]
\centering

\begin{tcolorbox}[
    width=\textwidth,
    colback=brown!5!white,
    colframe=black,
    boxrule=0.5pt,
    arc=0pt,
    left=7pt,
    right=7pt,
    top=9pt,
    bottom=9pt
]
\small
\setstretch{1.08}

\tcbox[
    colback=black,
    colframe=black,
    coltext=white,
    boxrule=0pt,
    arc=2pt,
    left=6pt,
    right=6pt,
    top=2pt,
    bottom=2pt
]{\texttt{\# Prompt for multi-turn audio-visual inconsistency judgment}}

\vspace{0.6em}

Please carefully watch this video, analyze its audio and visual content, and determine whether there is audio-visual inconsistency.

\vspace{0.45em}
\noindent\textbf{Output format}

Please return in the following exact format:
\begin{enumerate}[leftmargin=1.5em,itemsep=0.25em,topsep=0.35em]
    \item \textbf{Is there inconsistency:} (Yes/No) -- Judge independently based on video content.
    \item \textbf{Inconsistency category:} (If Yes, choose one from the following 8 categories. If No, fill \texttt{No}.)
    \item \textbf{Inconsistency point description:} (If inconsistency exists, describe the specific inconsistency between visual and audio; if No, fill \texttt{Audio and video are consistent}.)
\end{enumerate}

\vspace{0.45em}
\noindent\textbf{8 Inconsistency Categories (must choose from these)}

\vspace{0.25em}
\noindent\textbf{Class 1 (Active Speaker -- interview/dialogue, person speaking in video)}
\begin{itemize}[leftmargin=1.4em,itemsep=0.22em,topsep=0.3em]
    \item \texttt{TEMPORAL\_SHIFT}: Temporal offset, audio leads or lags behind video by 0.5--2s (e.g., the mouth moves but the sound is delayed by 0.5s).
    \item \texttt{LIP\_SYNC}: Lip-sync mismatch, TTS-generated voice does not match lip movement (e.g., the video shows a man speaking but the voice sounds like a woman).
    \item \texttt{VOICE\_IDENTITY}: Voice identity conflict, the speaker's voice changes abruptly (e.g., the video shows an elderly person but the voice is a child's).
    \item \texttt{VOLUME\_FLUCTUATION}: Volume conflict, the person is still but the volume fluctuates (e.g., the person stands still but the voice sounds as if it is moving closer and farther away).
\end{itemize}

\vspace{0.15em}
\noindent\textbf{Class 2 (Voiceover -- narration, no speaker in video)}
\begin{itemize}[leftmargin=1.4em,itemsep=0.22em,topsep=0.3em]
    \item \texttt{SEMANTIC\_DIVERGENCE}: Semantic inconsistency, TTS text contradicts video content (e.g., the video shows food but the narration talks about phones).
    \item \texttt{BACKGROUND\_CONFLICT}: Background sound conflict, narration or music contradicts the video scene (e.g., the video shows an office but the background contains bar music).
\end{itemize}

\vspace{0.15em}
\noindent\textbf{Class 3 (Scenic -- scenery/scene, no human voice)}
\begin{itemize}[leftmargin=1.4em,itemsep=0.22em,topsep=0.3em]
    \item \texttt{EMOTION\_MISMATCH}: Background music emotion mismatch, the video is a sad scene but has happy music (e.g., the video shows a funeral but the music is joyful).
    \item \texttt{BACKGROUND\_SOUND}: Background sound conflict, the video is scene A but the audio is scene B (e.g., the video shows a forest but has city traffic sounds; or the video shows chopping but no chopping sound).
\end{itemize}

\vspace{0.45em}
\noindent\textbf{Multi-turn questions}

\begin{enumerate}[leftmargin=1.5em,itemsep=0.25em,topsep=0.35em]
    \item \textbf{Question 1:} Is there audio-visual inconsistency? Answer \texttt{Yes} or \texttt{No} only.
    \item \textbf{Question 2:} What is the inconsistency category? Choose from: \texttt{EMOTION\_MISMATCH}, \texttt{BACKGROUND\_SOUND}, \texttt{SEMANTIC\_DIVERGENCE}, \texttt{LIP\_SYNC}, \texttt{TEMPORAL\_SHIFT}, \texttt{VOLUME\_FLUCTUATION}, \texttt{VOICE\_IDENTITY}, \texttt{BACKGROUND\_CONFLICT}. If consistent, answer \texttt{No}.
    \item \textbf{Question 3:} Please explain your reasoning process for the above judgment. Describe what you observed in the video and what you heard in the audio.
\end{enumerate}

\end{tcolorbox}

\caption{Prompt for multi-turn audio-visual inconsistency judgment.}
\label{fig:multiturn_av_inconsistency_prompt}
\end{figure*}



\section{Annotation Quality Analysis}\label{sec:appendix_annotation_quality}

To ensure the reliability and consistency of AVID annotations, we conducted inter-annotator agreement studies at two critical stages of our construction pipeline.

\subsection{Strategy Planning Stage}
At the strategy planning stage, we sampled 200 video segments and recruited 3 annotators to independently evaluate the feasibility decisions and injection type selections made by the strategy agent. Each annotator was asked to determine: (1) whether the selected segment is suitable for inconsistency injection (feasibility), and (2) whether the chosen injection type is semantically appropriate for the given segment class. We computed Cohen's Kappa coefficient to measure agreement among annotators.

The results show substantial agreement with a Cohen's Kappa of $\kappa = 0.78$, indicating that the strategy agent's decisions are largely aligned with human judgment. The remaining disagreements primarily occurred in edge cases where segment boundaries were ambiguous or where multiple injection types could be equally valid.

\subsection{Post-construction Quality Control Stage}
After video construction, we randomly sampled 300 constructed videos (150 inconsistent and 150 consistent) and asked 3 annotators to independently verify: (1) whether the video is truly inconsistent/consistent, (2) what is the inconsistency category (if applicable), and (3) whether the provided temporal boundaries and reasoning explanations are accurate.

The inter-annotator agreement for inconsistency detection is $\kappa = 0.85$, for category classification is $\kappa = 0.81$, and for temporal boundary accuracy (within 1 second tolerance) is $\kappa = 0.79$. These results demonstrate that AVID annotations achieve high consistency across human annotators, ensuring reliable evaluation protocols.

\end{document}